\documentclass[fleqn,usenatbib]{mnras}

\usepackage{newtxtext,newtxmath}

\usepackage[T1]{fontenc}
\usepackage{ae,aecompl}

\usepackage{graphicx}	
\usepackage{amsmath}	
\usepackage{hyperref}
\usepackage{float}
\usepackage[caption = false]{subfig}

\title[MUSIC and NIKA2 twin sample HE mass bias]{Exploring the hydrostatic mass bias in MUSIC clusters: application to the NIKA2 mock sample}

\author[G. Gianfagna et al.]{
Giulia Gianfagna,$^{1}$\thanks{E-mail: gianfagna.1665033@studenti.uniroma1.it }
Marco De Petris,$^{1}$
Gustavo Yepes,$^{2}$
Federico De Luca,$^{1,3}$
\newauthor Federico Sembolini,$^{1,2}$
Weiguang Cui,$^{4}$
Veronica Biffi,$^{5,6}$
Florian K\'eruzor\'e,$^{7}$
\newauthor Juan Mac\'ias-P\'erez,$^{7}$
Fr\'ed\'eric Mayet,$^{7}$
Laurence Perotto,$^{7}$
Elena Rasia,$^{6,8}$
\newauthor Florian Ruppin$^{9}$
\\
$^{1}$Dipartimento di Fisica, Sapienza Universit\'a di Roma, Piazzale Aldo Moro, 5-00185 Roma, Italy\\
$^{2}$Departamento de F\'{\i}sica Te\'orica and CIAFF, M\'odulo 8, Facultad de Ciencias, Universidad Aut\'onoma de Madrid, 28049 Madrid, Spain\\
$^{3}$Dipartimento di Fisica, Universit\'a di Roma `Tor Vergata', Via della Ricerca Scientifica, I-00133 Roma, Italy\\
$^{4}$Institute for Astronomy, University of Edinburgh, Royal Observatory, Edinburgh EH9 3HJ, UK\\
$^{5}$Universit\"{a}ts-Sternwarte M\"{u}nchen, Fakult\"{a}t f\"{u}r Physik, LMU Munich, Scheinerstr. 1, 81679 M\"{u}nchen, Germany \\
$^{6}$IFPU - Institute for Fundamental Physics of the Universe, Via Beirut 2, 34014 Trieste, Italy \\
$^{7}$Univ. Grenoble Alpes, CNRS, Grenoble INP, LPSC-IN2P3, 53, avenue des Martyrs, 38000 Grenoble, France\\
$^{8}$INAF Osservatorio Astronomico di Trieste, via Tiepolo 11, I-34131, Trieste, Italy\\
$^{9}$Kavli Institute for Astrophysics and Space Research, Massachusetts Institute of Technology, 77 Massachusetts Avenue, Cambridge, MA 02139 
}

\date{Accepted XXX. Received YYY; in original form ZZZ}

\pubyear{2020}

\begin{document}
\label{firstpage}
\maketitle

\begin{abstract}
Clusters of galaxies are useful tools to constrain cosmological parameters, only if their masses can be correctly inferred from observations. In particular, X-ray and Sunyaev-Zeldovich (SZ) effect observations can be used to derive masses within the framework of the hydrostatic equilibrium. 
Therefore, it is crucial to have a good control of the possible mass biases that can be introduced when this hypothesis is not valid. 
In this work, we analyzed a set of 260 synthetic clusters from the MUSIC simulation project, at redshifts $0 \leq z \leq 0.82$. We estimate the hydrostatic mass of the MUSIC clusters from X-ray only (temperature and density) and from X-ray and SZ (density and pressure). Then, we compare them with the true 3D dynamical mass. The biases are of the order of 20\%. 
We find that using the temperature instead of the pressure leads to a smaller bias, although the two values are compatible within 1$\sigma$. Non-thermal contributions to the total pressure support, arising from bulk motion and turbulence of the gas, are also computed and show that they are sufficient to account for this bias. 
We also present a study of the correlation between the mass bias and the dynamical state of the clusters. A clear correlation is shown between the relaxation state of the clusters and the bias factor. We applied the same analysis on a subsample of 32 objects, already selected for supporting the NIKA2 SZ Large Program.
\end{abstract}

\begin{keywords}
methods: numerical -- galaxies: clusters: general -- galaxies: clusters: intracluster medium -- cosmology: large-scale structure of Universe
\end{keywords}



\section{Introduction}

Galaxy clusters are the most massive gravitationally bound objects in the Universe and they are mainly composed by Dark Matter, that amounts to 80 \% of the total mass (for a full review see \citealt{bib:rev}). About 8\% is composed by galaxies and the remaining 12\% is represented by the so called Intra Cluster Medium (ICM), i.e. the hot gas located between galaxies. This gas component provides significant physical information, as it can be observed in the X-ray band and in millimeter wavelengths through the Sunyaev-Zeldovich effect \citep{bib:z1}.

The emission in the X-ray band is mainly due to the thermal bremsstrahlung. From this emission we can directly measure the temperature, which determines the bremstrahlung cut-off, and the electron density, to which the spectrum normalization is proportional (for a review see \citealt{Boehringer}). X-ray observations occurred to be particularly successful because the emission is proportional to the square of the gas density (for a review on the methods adopted to reconstruct the mass profiles in X-ray luminous galaxy clusters see \citealt{bib:rev_x}). In the last two decades, X-ray observatories with improved sensitivity and angular resolution, like \textit{XMM-Newton} and \textit{Chandra}, has conducted cluster X-ray emission studies over large areas of the sky \citep{bib:Pacaud2007, andreon}, and deeper studies of previously known objects (e.g. \citealt{bib:Vikhlinin2009, bib:Mantz2010}).

X-ray observations are mainly exploring the central regions of the clusters, where the electron number density is high, although there are also X-ray projects exploring the cluster outskirts, such as X-COP \citep{bib:xcop}. A more efficient way to map the cluster outskirts is through the thermal Sunyaev-Zeldovich effect (SZ, \citealt{bib:z1}). This effect depends linearly on the pressure, so it is less sensitive to the density decrease at high radial distances from the cluster centre. The SZ effect is a redshift independent probe that consists of the spectral distortion of the CMB radiation due to the inverse Compton scattering of the CMB photons with the hot electrons of the ICM (for a review see \citealt{bib:sz2, bib:dn, Mroczkowski2019}). It can be used to directly measure the ICM pressure distribution. The latter can be combined with the electron density from an X-ray observation to infer the cluster mass profile. This method avoids the necessity of obtaining the  deep X-ray observations required to measure a spatially resolved temperature profile. To date, the SZ effect induced by thermal electrons (tSZ) has been detected for more than a thousand galaxy clusters, including more than 200 new clusters previously unknown by any other observational means \citep{bib:dn} thanks to new observing facilities such as the South Pole Telescope (SPT) \citep{bib:Williamson2011, bib:Reichardt2013, Bleem2020}, the Atacama Cosmology Telescope (ACT) \citep{ bib:Marriage2010, bib:Hasselfield2013, Hilton2018, hilton2020}, and the \textit{Planck} satellite \citep{ bib:PlanckCollaboration2011, bib:PlanckCollaboration2013, bib:Planck}.

Both X-ray and SZ observations can be used to infer the mass of a cluster. In particular, from the former we can exploit the temperature and the electron density, from the latter the pressure profile. In order to use this informations, we need to make three fundamental assumptions: the gas must trace the cluster potential well, it must be spherically symmetric and in hydrostatic equilibrium (HE, \citealt{bib:rev} for a review). The mass inferred through this method is called hydrostatic mass (see Section \ref{sec:HE}). An easy way to track the error made when using HE is the hydrostatic mass bias $b$ \citep{bib:planck_2013}. It is defined as the difference of the cluster total mass to the one estimated by hydrostatic equilibrium, divided by the total mass $b = (M_{\rm tot} - M_{\rm HE})/M_{\rm tot}$.

Galaxy clusters, and, in particular, their number counts are a fundamental cosmological tool. The abundance of clusters and its evolution with redshift are particularly sensitive to the cosmic matter density, $\Omega_{\rm m}$, and the present amplitude of density fluctuations, characterized by $\sigma_8$, i.e. the rms linear overdensity in spheres of radius $8h^{-1}$ Mpc. The CMB primary anisotropies, on the other hand, are related to the density perturbation power spectrum at the time of recombination. A comparison of the amplitude of density perturbations from recombination until today, allows us to look for possible extensions to the concordance $\Lambda$CDM model, such as non-minimal neutrino masses or non-zero curvature contributions \citep{bib:planck_2015, bib:salvati}.

In \citet{bib:planck_2013}, a tension between the number of clusters detected by SZ signal, and the number of clusters predicted from the cosmological parameters inferred from the primary CMB power spectrum is reported. This issue was later confirmed in \citet{bib:planck_2015} and arises when the HE mass bias is fixed to a constant value of 0.2 in cluster cosmological analyses.
The hydrostatic mass bias $b$ plays an important role in the number counts, because it leads to a modification of the cluster population at a given mass. In particular, it significantly affects the value of $\sigma_8$: the lower $(1 - b)$ is, the higher is $\sigma_8$, \citep{bib:salvati, ruppin2019_p}. \citet{bib:salvati} published an update of the constraints on cosmological parameters from the clusters observed by \textit{Planck}. They find that the bias needed to reconcile CMB constraints with those from the tSZ number counts is $(1-b) = 0.62 \pm 0.07$, which is compatible with the value $(1-b) = 0.58 \pm 0.04$ found in \citet{bib:planck_2015}. This value is confirmed by \citet{Koukoufilippas2020}, who cross-correlate \textit{Planck} maps of the tSZ Compton-y parameter with the galaxy distribution.

These values derived from observations are nevertheless in disagreement with the value of $(1 - b)$ estimated from simulations, which is about $0.8$. This topic will be deeply discussed later in this work. Moreover, there is a factor of $2.5$ more clusters predicted than observed when taking into account the CMB cosmology and a value $(1-b)$ of 0.8 \citep{andreon_2, bib:planck_2015, bib:salvati}. This value of the bias, derived from simulations, has been recently confirmed by \citet{bib:Makiya2020}. They perform a joint analysis of power spectra of the tSZ and the cosmic weak lensing shear, in order to obtain a $(1-b)$ constraint which is independent from the primordial CMB spectrum. They find $(1-b) = 0.73^{+0.08}_{-0.13}$, and conclude that the late-time probes (tSZ and cosmic weak lensing shear) cosmologies are consistent with each other, but they could not be totally consistent with the CMB cosmology, which is leading to a different value of the mass bias.

However, the cluster number counts are limited by systematic effects, in particular those affecting the mass estimates. The tSZ power spectrum, in turn, is not measured with sufficient accuracy, especially at small angular scales, to reduce the tension with the CMB. The tSZ cosmological analysis can be improved by considering more realistic and complex hypotheses on the mass bias (e.g. redshift and/or mass dependence), the pressure profile and mass function \citep{bib:salvati, ruppin2019_p}.

The tension between the cluster number counts from the tSZ and the CMB power spectrum arising from fixing the hydrostatic bias to 0.2 in cosmological analysis has led the scientific community to investigate this discrepancy in more detail. Several studies have been made on the HE mass bias. In this work, a simulated dataset of almost 260 clusters from the MUSIC simulations is analysed. We focus on the determination of cluster masses assuming the hydrostatic equilibrium hypothesis, at different redshifts, and compare them with the true cluster mass derived from the simulation data. The HE masses can be estimated using two different equations \citep{bib:pratt}, depending on which ICM thermodynamic quantities are used, i.e. the pressure and electron density (hereafter referred to as SZ mass) or the temperature and electron density (hereafter X-ray mass). There are also different ways of computing the radial gradients of these quantities. In this work, we present a complete study of the HE mass derivations from the different assumptions. Moreover, we also correlate the results for the mass bias with the dynamical state of the MUSIC clusters. A correction to the HE mass, which takes into account the non thermal pressure contribution arising from gas motions in the ICM, is applied. 

We repeated the same analysis for a sub sample of 32 objects from MUSIC, in the redshift range $0.5<z<0.9$. This sample, named the NIKA2 twin sample, has been selected to be representative of the clusters observed in the NIKA2 SZ Large Program (\citealt{mayet}, LPSZ). The LPSZ uniquely exploits the excellent match in sensitivity and spatial resolution of \textit{XMM-Newton} and the NIKA2 camera, which is a millimetre camera installed at the 30-m radio telescope of the Institut de Radioastronomie Millim\'etrique (IRAM) in Pico Veleta, Spain \citep{bib:NIKA2, perotto}. A previous analysis of the gas pressure profiles reconstruction on this sample by applying the NIKA2 data reduction pipeline has been already presented in \citet{bib:ruppin}.

The paper is organized as follows. In Section \ref{sec:HE}, we describe the hydrostatic equilibrium, and the estimate of the cluster mass from different approaches. In Section \ref{sec:massbias_biblio}, we review the estimates of the HE mass bias from different numerical hydrodynamical simulations that have been published in literature, showing the relatively large variations in the bias results from the different simulation suites. In Section \ref{sec:sim_dataset}, we briefly introduce the MUSIC simulations and the data used in this analysis, classifying clusters by their dynamical state. The ICM profiles are presented in Section \ref{sec:profiles}. The distribution of the results for HE masses and their biases are described in Section \ref{sec:res}, along with the modelling of the non-thermal correction. In Section \ref{sec:nika2} and Appendix \ref{app:nika2} we focus on the HE masses and biases for the NIKA2 twin sample. Finally, in Section \ref{sec:conclusion}, the main conclusions from these analyses are given.

\section{Hydrostatic equilibrium}
\label{sec:HE}

We use the hydrostatic equilibrium (HE) hypothesis to estimate the mass of each clusters, see \cite{bib:rev} for a review. It assumes that the gas thermal pressure is balanced by the gravitational force, so that the cluster is in equilibrium. Further assuming that the system is in spherical symmetry and the gas pressure is purely thermal, the total mass inside a sphere of radius $r$, can be written as
\begin{equation}
M_{\rm HE}(r) = -\frac{r^2}{G \mu m_p n_{\rm e}(r)} \frac{\rm d P_{\rm th}(r)}{\rm dr}
\label{eq:Mhe_P}
\end{equation}
where $G$ is the gravitational constant, $\mu$ is the mean molecular weight of the ICM, here 0.59, $m_p$ is the proton mass, $n_{\rm e}$ and $P_{\rm th}$ are the numerical electron density and the thermal pressure of the gas. Assuming the equation of state of an ideal gas, it follows that the cluster mass can also be derived from the electron density and temperature $T(r)$ profiles, as
\begin{equation}
M_{ \rm HE}(r) = - \frac{rk_{\rm B} T(r)}{G\mu m_p}\left[ \frac{\rm d\ln n_{\rm e} (r)}{\rm d\ln r} + \frac{\rm d\ln T(r)}{\rm d\ln r}\right].
\label{eq:Mhe_T}
\end{equation}
We refer to Eq.(\ref{eq:Mhe_P}) as $M_{\rm HE, SZ}$ given that the pressure is estimated by SZ observations, while we refer to Eq.(\ref{eq:Mhe_T}) as $M_{\rm HE, X}$ because the temperature and the electron density are usually estimated from X-rays observations \citep{bib:rasia19}.

Deviations from equilibrium could have an impact on observable properties of clusters and may cause systematic errors when Eq.(\ref{eq:Mhe_P}) and (\ref{eq:Mhe_T}) are used to estimate cluster masses. A direct comparison with the real mass is usually quantified in the form of a hydrostatic mass bias $b$ (see references in Section \ref{sec:massbias_biblio}). It should be stressed that the mass bias can be estimated only if the exact cluster total mass is available, which is the case in simulations. The real mass $M_{\rm true}$ of a simulated cluster can be easily computed by summing all the dark matter, stars and gas particle/cells masses inside an aperture radius. The mass bias $b_{\rm SZ}$ or $b_{\rm X}$, at a specific radius, is defined as

\begin{equation}
\label{eq:bias}
b = \frac{M_{\rm true} - M_{\rm HE}}{M_{\rm true}}.
\end{equation}
The bias defined in this way is usually a positive quantity, since the HE mass often underestimates the true mass. It can happen also the contrary, leading to a negative bias. Sometimes in literature the opposite difference between the masses is chosen.

\section{Hydrostatic mass bias, state of the art}
\label{sec:massbias_biblio}

The comparison between tSZ cluster number counts and CMB \textit{Planck} results has led the community to carefully account for the impact of the HE mass bias on cosmological constraints. Here we focus on the HE mass bias $b$, Eq.(\ref{eq:bias}), computed at $R_{500}$\footnote{The radius where the cluster density is 500 times the Universe critical density $\rho_{\rm c}$ at that time, $\rho_{\rm c} = 3 H(a)^2 / (8 \pi G)$ where $H(a)$ is the Hubble function. $M_{500}$ is the mass inside a sphere with radius $R_{500}$.}. This parameter has been extensively studied with a variety of numerical hydrodynamical simulations. 

\subsection{Previous work in literature}

\begin{figure}
 \centering
 \includegraphics[width=0.5\textwidth]{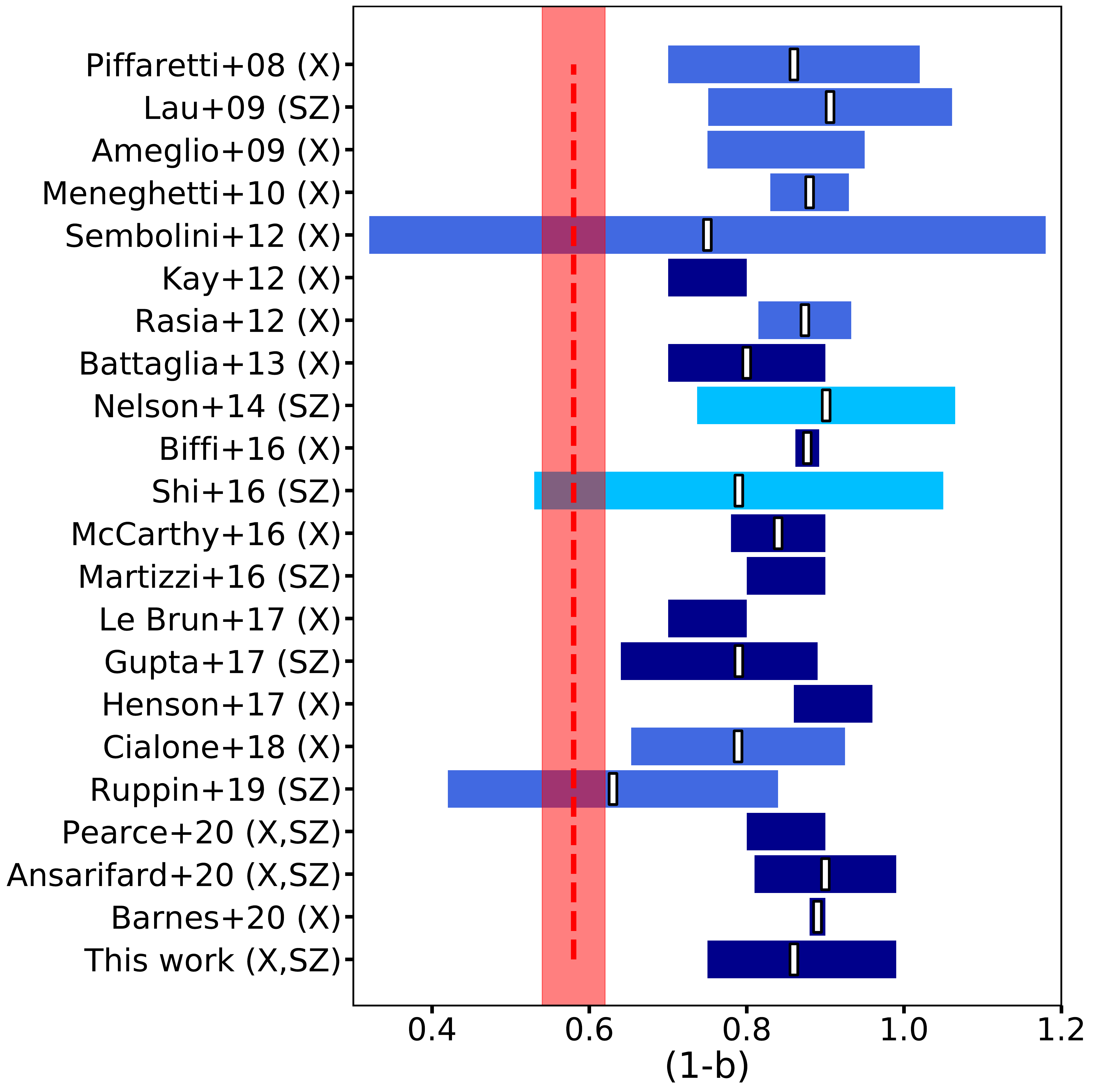}
 \caption{ A compilation from literature of the $(1-b)$ mean values (white rectangle) and their errors (blue bars). They are derived from different hydrodynamical simulation works using the mass-weighted temperature profiles. The vertical red dashed line and its shaded area are the value and the error $(1-b) = 0.58 \pm 0.04$, needed to reconcile the cosmological constraints obtained from \textit{Planck} cluster counts and CMB power spectrum \citep{bib:planck_2013, bib:planck_2015}. The three different shades of blue represent the physics included in each simulations, from the basic set of physical processes, Non radiative in light blue, cooling+star formation+supernovae feedback, in medium blue, to the complete set, including the effects of super massive black hole feedbacks in dark blue. In the vertical axis, we show the reference for each work and in parenthesis, the type of bias that is derived, from X-Ray, SZ or both.}
 \label{fig:sims_b}
\end{figure}

In Fig. \ref{fig:sims_b} we present a compilation of results published in the literature, including the error estimates of the mean values for $1-b$. The mean values are represented as vertical white rectangles, the errors corresponding to each value are the blue coloured regions. The different shades of blue represent the physical processes included in each simulation: non-radiative (light blue), cooling+star formation and supernovae feedback (medium blue) and those including also Super Massive Black Hole Feedback (SMBH) (dark blue). In the vertical axis, after the authors reference, we indicate, in parenthesis, the type of bias estimated in each work: SZ for HE masses derived from pressure and density profiles and X-ray for HE masses computed from temperature and density profiles, see Section \ref{sec:HE}. 

Some authors define a positive bias, see the definition in Eq. (\ref{eq:bias}), others negative, but in Fig. \ref{fig:sims_b}, in order to compare them, all biases are taken as positive. Here we represent the biases from the mass-weighted temperature profiles, in order to focus mainly on the degree of HE bias, since the spectroscopic-like temperature profiles \citep{mazzotta2004} are sensitive, in addition, to observational biases in the derived gas density and temperature profiles.

As it can be seen in the Figure, most of the published results for $1-b$ are in the range $0.75$ to $0.9$, or $b \sim 0.25-0.10$. The majority of these results are in disagreement with respect to the bias needed to reconcile CMB constraints with those from the tSZ number counts of $(1-b) = 0.58 \pm 0.04$, shown as the vertical red dashed line and shaded region \citep{bib:planck_2013, bib:planck_2015}. Only 3 bias values over 22 are compatible with it within $1\sigma$.

The mean and standard deviation are reported for all the works except for \citet{bib:biffi}, where median and MAD (Median Absolute Deviation, computed as $median(|b_i - median(b)|) / n_{\rm clusters}$) are presented. We find a X-ray bias of $b_{\rm X}=0.14^{+0.11}_{-0.13}$, using the median and the percentiles (16th and 84th). Also \citet{bib:gupta} report median and percentiles, while \citet{barnes2020} report the bootstrap errors. There are also some authors, represented without the vertical white rectangle, who give a range of values for $b$ but not a central value with an error \citep{bib:ameglio, bib:kay, bib:martizzi, bib:lebrun, Henson2017, bib:pearce}.

\subsection{Bias dependence on simulation and sample properties}

\begin{figure}
 \centering
 \includegraphics[width=0.5\textwidth]{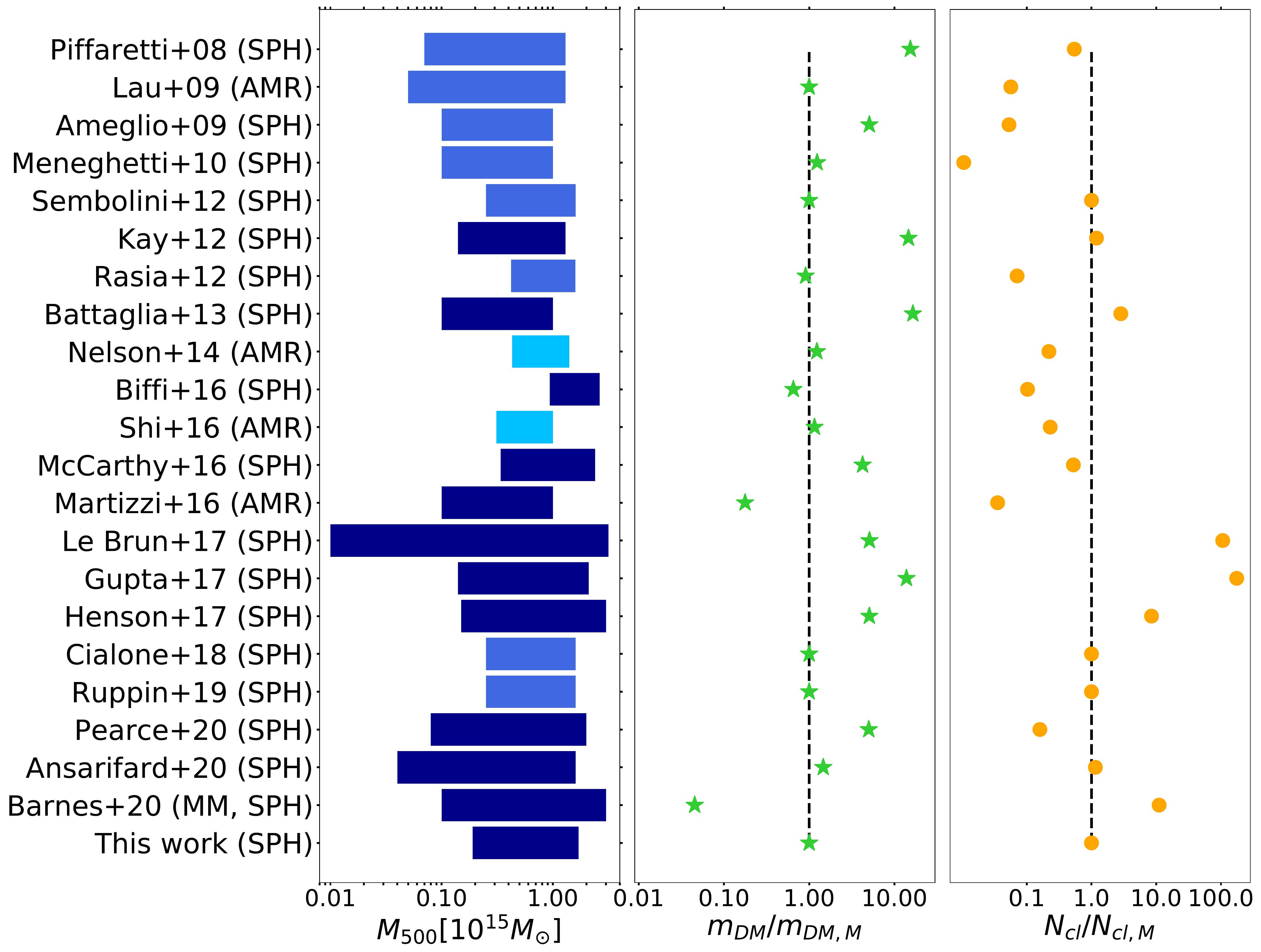}
 \caption{Comparison of the main properties of the simulation works.
 Left panel: The $M_{500}$ mass range of clusters used for each simulation work. The color of the bars have the same meaning as in Fig. \ref{fig:sims_b}. \protect\citet{bib:meneghetti, bib:martizzi} do not give the exact $M_{500}$ range, so we give an estimation, from $10^{14}$ to $10^{15}$. Central panel: the mass resolution of Dark Matter particles, relative to the MUSIC DM resolution $m_{\rm DM,M} = 1.3\times10^9 \rm M_{\odot}$, used in this work. Right panel: the number of cluster objects used in each publication, relative to the number of clusters used in this work ($N_{\rm cl,M}=282)$. In the vertical axis, next to the reference of each work, we show in parenthesis the computational hydrodynamical method used by each simulation: SPH or AMR. \protect\citet{barnes2020} employ three different simulations: Illustris (Moving Mesh), BAHAMAS (SPH) and MACSIS (SPH), in this Figure the DM mass from Illustris is represented, since the MACSIS and BAHAMAS codes are also used in \protect\citet{Henson2017} and have the same DM mass. }
 \label{fig:sims_m}
\end{figure}

This compilation shows a wide spread in the determination of the bias parameter and their errors. It is obvious that these differences might be attributed to the particularities of the simulations used in each work. To shed some light on this problem, we also account for features like mass resolutions (Fig. \ref{fig:sims_m}, central panel); the range of analyzed halo masses (Fig. \ref{fig:sims_m}, left panel) and the statistics of the total number of clusters (Fig. \ref{fig:sims_m}, right panel). Moreover, the physical processes included in each simulation should also be considered. They are represented in Fig. \ref{fig:sims_b} and Fig. \ref{fig:sims_m} by the different shades of blue of the bars, as already explained above. 

In Fig. \ref{fig:sims_m}, we indicate the type of code used in each simulation in parenthesis next to the reference of the study. The majority of them are based on different flavours of the Smoothed-Particle Hydrodynamics (SPH). \citet{bib:lau, bib:nelson, bib:shi, bib:martizzi} use finite volume eulerian hydrodynamics with Adaptive Mesh Refinement algorithms.
The Illustris simulations \citep{barnes2020}, using the Moving Mesh code (MM) AREPO \citep{arepo} have the lowest DM particle mass, along with the RAMSES code used by \citet{bib:martizzi}. \citet{bib:gupta} with Magneticum simulations, and \citet{bib:lebrun} with cosmo-OWLS, employ the largest number of clusters and have one of the lowest errors on the bias estimate.

We compared the HE biases in Fig. \ref{fig:sims_b} with the main features of each simulation shown in Fig. \ref{fig:sims_m}. We further plot the biases as a function of each quantity. For the sake of brevity, we do not show these figures, but only conclude here that we cannot draw any clear dependence between the HE bias and the cluster mass range (left panel of Fig. \ref{fig:sims_m}), the particle mass resolution (central panel) or the number of clusters included in the analysis (right panel). 

The simulation physics comprehends a wide range of processes, but here we gather them in three classes: those which includes the gravitational and non radiative physics (NR, light blue in Fig. \ref{fig:sims_b}); the `middle' set, which includes also radiative processes, like cooling, star formation and Supernovae feedback (CSF, medium blue); and the `complete' set, which, in addition, includes also the feedbacks from super massive black holes (CSF+SMBH, dark blue). The simulations which have the SMBH feedback seem to have lower errors on the bias, such as \citet{bib:kay, bib:battaglia, bib:biffi, bib:mccarthy, bib:martizzi, bib:gupta, bib:pearce, bib:rasia19, barnes2020}. Even in \cite{bib:meneghetti} the bias error is low mostly for two main reasons: their simulations considered 1/3 of Spitzer thermal conductivity which homogenize the medium and the statistics are  limited due to the small sample size (see Fig. \ref{fig:sims_m}). It is not straightforward  to derive the impact of SMBH on the HE mass bias from the different simulation results  compiled in  Fig. \ref{fig:sims_b}, because each simulation has different features, see Fig. \ref{fig:sims_m}. In this paper, we are comparing an homogeneous and statistical significant set of clusters simulated with different physics flavours. Thus, we can provide a definitive answer on this issue (see Section \ref{sec:baryon_models}).

\subsection{Bias dependence on other quantities or measurements}

The more compelling questions about the HE mass bias are its dependence on redshift, on cluster dynamical state or on whether the bias is calculated from spectroscopic-like temperature. The redshift dependence has been studied by several authors \citep{bib:piffaretti, bib:lau, bib:lebrun, Henson2017}. The expectation is that clusters, going at higher redshifts and being less relaxed, tend to have more mass in substructures \citep{neto, angelinelli2019} risking to violate the hydrostatic equilibrium hypothesis. However, this behaviour has not yet been confirmed in any work. 

According to their dynamical state, clusters are usually classified in two main classes: the relaxed ones, well described by spherical symmetry and HE, and disturbed clusters, the opposite. Non-thermal gas motions in the ICM and the non-spherical symmetry of a cluster (usually disturbed) will most likely lead to a larger HE bias and a larger scatter, with respect to the more regular (relaxed) clusters. Several authors analysed the mass bias dependence with the dynamical state, like \citet{bib:piffaretti, bib:rasia, bib:nelson, Henson2017, bib:rasia19}. They all find no significant distinction between the mass bias of regular and disturbed clusters, given the large dispersion \citep{bib:cialone}. However, \citet{bib:biffi} observe that Cool-Core (CC) and Non Cool-Core (NCC) clusters behave differently, with a larger bias for NCC, especially in the innermost cluster regions.

The spectroscopic-like temperature, see Section \ref{sec:T}, is estimated by weighting the temperature by the X-ray emission of each gas particle. Usually the HE mass bias estimated from this temperature by combining it with the electron density is larger than the one from the mass-weighted temperature and shows a dependence with the true mass of the halos. This is mainly due to temperature inhomogeneities \citep{rasia06, bib:rasia, rasia14, bib:lebrun, Henson2017, bib:pearce, barnes2020}. Moreover, \citet{bib:piffaretti} find that the spectroscopic bias also depends on the dynamical state of clusters, with large biases found in the most disturbed clusters (see also \citealt{Biffi2014}).

In the last years, the HE mass biases (SZ and X-ray) were often studied together \citep{bib:pearce, bib:rasia19}. \cite{bib:rasia19}, for example, analyse more than 300 simulated massive clusters, from `The Three Hundred Project' \citep{Cui2018}. They find that a robust correction to the HE mass bias can be inferred when the gas inhomogeneity from X-ray maps are combined with the asymptotic external slope of the gas density or pressure profiles, which can be derived from X-ray and SZ effect observations. Both SZ and X-ray biases are estimated, with values of 10\%, by using models to fit ICM radial profiles.

\section{The simulated dataset}
\label{sec:sim_dataset}

\subsection{MUSIC simulations}
The clusters analysed in this work are taken from the MUSIC project \citep{bib:semb} which consists of two sets of resimulated clusters extracted from two large volume simulations: the MUSIC-1 sample, extracted from the $500h^{-1}$ Mpc MareNostrum Universe simulation box \citep{bib:29}, and the MUSIC-2 sample, from the $1h^{-1}$ Gpc MultiDark (MD) simulation box \citep{bib:30}.

In this work, the 258 zoomed regions around the most massive clusters in the MUSIC-2 database were analysed. From a low resolution version ($256^3$ particles) of the two simulations, the particles inside a sphere of $6h^{-1}$ Mpc radius at $z = 0$ are mapped back to the initial redshift, using the \citet{bib:32} zooming technique, to identify their corresponding Lagrangian regions. These regions are then resimulated with high resolution and populated with SPH gas particles. The original MD dark-matter-only simulation was performed with L-Gadget2 code \citep{klypin2016} and adopting WMAP7+BAO+SNI cosmology: $\Omega_{\rm M} = 0.27$, $\Omega_{\rm b} = 0.0469$, $\Omega_{\Lambda} = 0.73$, $\sigma_8 = 0.82$, $n = 0.95$ and $h = 0.7$ \citep{bib:WMAP}.

All the resimulations are done with the TreePM+SPH GADGET code \citep{bib:gadget2}, and include three different classes of physical processes, labelled as \textit{flavours}: Non Radiative (NR), Cooling and Star Formation (CSF) and Active Galactic Nuclei (AGN). The NR flavour includes only the gravitational and gasdynamical effects, while the CSF flavour includes radiative processes, like star formation, feedback from supernovae, UV photoionization, and radiative cooling \citep{bib:semb}, and finally AGN, where the AGNs and their feedback are added, using the models for super massive black hole feedbacks \citep{planelles}.

The clusters were identified using a Bound Density Maxima halo finder (see also AHF halo finder, \citealt{Knollmann_2009}). Since we take all the objects above a given mass, the MUSIC-2 catalogue constitutes a complete volume limited sample. Our cluster masses $M_{500}$ range from $1.9\times 10^{14} \rm M_{\odot}$ to $1.7 \times 10^{15} \rm M_{\odot}$ at $z = 0$. All of them were resimulated with NR, CSF and AGN flavours with a DM and gas mass resolution of $m_{\rm DM} = 1.29\times10^9 \rm M_{\odot}$ and $m_{\rm gas} = 2.7 \times 10^8 \rm M_{\odot}$. MUSIC-2 cluster regions have been saved at specific redshifts, so in this analysis we also studied the cosmic evolution of these objects, using only the main progenitors of the $z=0$ clusters at redshifts 0.11, 0.33, 0.43, 0.54, 0.67 and 0.82. 

\subsection{The NIKA2 LPSZ twin sample}
\label{sec:nika2_intro}

From the MUSIC-2 dataset a sub sample of objects has been selected to reproduce the clusters present in the NIKA2 Large Program SZ (LPSZ) catalogue \citep{mayet}. NIKA2 \citep{bib:NIKA2, calvo, perotto}, is the new multipixels camera at 150 and 260 GHz installed at the 30-m telescope of the Institut de Radioastronomie Millim\'etrique (IRAM). The NIKA2 SZ large program consists of mapping the tSZ signal of a representative sample of 50 galaxy clusters at high angular resolution ($18''$ and $11''$ in two bands) and in the $0.5 < z < 0.9$ redshift range. The cluster sample was extracted from the tSZ catalogues established by the \textit{Planck} and ACT collaborations \citep{bib:Planck, bib:Hasselfield2013}, and the selected clusters homogeneously populate the mass range with $M_{500} > 3 \times 10^{14} \rm M_{\odot}$ \citep{mayet} and redshift range. The MUSIC NIKA2 twin sample closely matches the same mass-redshift space as the NIKA2 tSZ large program. For the redshift bin $0.5 < z < 0.7$ eighteen clusters were chosen from the MUSIC-2 catalogue at redshift $0.54$. For the $0.7 < z < 0.9$ bin, 14 clusters from MUSIC redshift 0.82 were also selected. The same mass cut applied to the \textit{Planck} and ACT catalogue of $M_{500} > 3 \times 10^{14} \rm M_{\odot}$ was applied to the MUSIC sample, so only the clusters with a HE mass estimate above this threshold are used, in order to make the two sample comparable. A previous analysis on the gas pressure profiles recovered from the NIKA2 twin sample has been already performed \citep{bib:ruppin, bib:ppnika}.

\subsection{Characterization of the cluster dynamical state and morphology}
\label{subsec:dyn_state}

Throughout this analysis the dynamical state of the clusters has been inferred by two 3D indicators (\citealt{neto, bib:semb, bib:cialone}, for a novel approach to infer the cluster morphology and dynamical state with the Zernike polinomials see \citealt{valentina}). Here we focus on the dynamical state of clusters inside $R_{500}$, which is more in agreement with what is measured in observations.
The considered 3D dynamical state estimators are

\begin{itemize}
\item $M_{\rm sub}/M_{500}$, the ratio between the mass of the most massive cluster substructure and the total cluster mass inside $R_{500}$. This indicator is mainly sensible to strong mergers. Therefore it is useful to find the really disturbed clusters. An alternative definition of $M_{\rm sub}$ would be to account for the mass of all the substructures within the aperture radius. This approach is more sensible to a cluster relaxation state \citep{bib:cui, federico};\\

\item $\Delta_r$, the offset between the central density peak, $\textbf{r}_{\delta} $, and the centre of mass of the cluster, $\textbf{r}_{\rm cm}$, normalized to the aperture radius $R_{500}$:
\begin{equation}
\Delta_{\rm r} = \frac{|\textbf{r}_{\delta} - \textbf{r}_{\rm cm}|}{R_{500}}.
\label{eq:delta_r}
\end{equation}
\end{itemize}

In order to have a relaxed cluster, both indicators should be smaller than a given threshold, which varies depending on the authors \citep{bib:Maccio2007, bib:DOnghia2007}. Here, following \citet{bib:cialone}, both indicators should be lower than 0.1 to define a cluster as relaxed, and greater than 0.1 to have a disturbed one. In the cases in which the two indicators provide contradictory answers, the cluster is classified as intermediate or hybrid. According to this selection for the dynamical state estimators, as seen in Table \ref{Tab:dyn_state}, the MUSIC dataset contains roughly 50\% of relaxed clusters for all the redshift intervals considered.
The same happens for the NIKA2 sub sample. 

\begin{table}
\centering
\caption{Fractions of relaxed, disturbed and intermediate clusters at each redshift for the MUSIC sample and for the NIKA2 sub sample.}
\begin{tabular}[t]{|c|c|c|c|c|c|c|}
\hline
 & \multicolumn{3}{c}{MUSIC} & \multicolumn{3}{c}{NIKA2} \\
\hline
 $z$  & Rel  & Interm & Dist  & Rel  & Interm & Dist \\
\hline
 0.00  & 52\% & 29\%   & 19\% &  &  & \\ 
 0.11 & 50\% & 34\%   & 16\% &  &  & \\ 
 0.33 & 53\% & 28\%   & 19\% &  &  & \\ 
 0.43 & 49\% & 34\%   & 17\% &  &  & \\ 
 0.54 & 47\% & 30\%   & 23\% & 44\% & 39\% & 17\% \\ 
 0.67 & 48\% & 30\%   & 22\% &  &  & \\ 
 0.82 & 52\% & 30\%   & 18\% & 57\% & 22\% & 21\% \\ 
\hline
\end{tabular}
\label{Tab:dyn_state}
\end{table}

We can combine the information provided by the two dynamical indicators in a single `relaxation' parameter $\chi_{DS}$, defined as in \citet{Haggar2020}, but dropping the virial ratio $\eta$, in order to keep the same number of relaxed clusters from the definition of the two separate dynamical indicators, $M_{sub}/M_{500}$ and $\Delta_r$ (see De Luca et al., in prep). $\chi_{DS}$ is a continuous, non binary, estimate of the dynamical state
\begin{equation}
    \chi_{\rm DS} = \sqrt{\frac{2}{\left(\frac{\Delta_{\rm r}}{0.1}\right)^2 + \left( \frac{M_{\rm sub}/M_{500}}{0.1}\right)^2 
    }} .
\end{equation}
All clusters that have this parameter higher than 1 are dynamically relaxed. We will study how this parameter is correlated with $b$ (in Section \ref{sec:morph_param_num}).

\section{ICM profiles of the MUSIC clusters}
\label{sec:profiles}

For each MUSIC cluster we compute the 3D radial profiles of the ICM thermodynamic properties. The cluster is divided into spherical shells, from 0 (the core) to $3R_{\rm vir}$ (the outskirts), each with a thickness of 10 kpc. The gas pressure and the electron density, are evaluated as the median of all the SPH gas particles inside each spherical shell, while the temperature used is the mass-weighted one. The associated uncertainty of the median value is estimated by \textit{MAD}
\begin{equation}
MAD = \rm median(|X_i - median(X)|)
\label{eq:mad}
\end{equation} 
The uncertainties associated to the median profiles usually increase in the cluster outskirts due to the deviations from the spherical symmetry and from a homogenous distribution of the ICM density and temperature (e.g. presence of clumps or disturbances generated by accreting material). 

\subsection{Pressure profile}
\label{subsec:Pprof}

The cluster pressure profile is well modelled by the generalized Navarro-Frenk-White (gNFW) model, introduced by \citet{bib:nagai}
\begin{equation}
\frac{P(r)}{P_{500}} = \frac{P_0}{x^{c} (1 + x^{a})^{\frac{b - c}{a}}}
\label{eq:gNFW}
\end{equation}
with $x = r/r_s$ a dimensionless radial distance normalised to the scale radius $r_s = R_{500}/c_{500}$, where $c_{500}$ is the concentration parameter. The parameters $b$ and $c$ are the slopes for outer and inner region radii respectively and $a$ is the steepness of the transition between the two regimes. The normalization $P_{500}$ is inferred by the scaling relation between the pressure content and the cluster total mass in the self-similar model (\citealt{bib:arnaud}, hereafter A10) purely based on gravitation:
\begin{equation}
P_{500} = 1.65 \times 10^{-3} E_{\rm z}^{8/3} \left[ \frac{M_{500}}{3 \times 10^{14} h_{70}^{-1} \rm M_{\odot} } \right]^{2/3} h_{70}^{2} \rm keV/cm^{3},
\label{eq:P500}
\end{equation}
$E_z$ is the ratio of the Hubble constant at redshift $z$ to its present value $H_0$, and $h_{70} = H_0 /$[70 km/s/Mpc]. 

A worthwhile step is to compare our simulated cluster sample with observations. A10 computed the mean pressure profile of galaxy clusters using observed clusters from REXCESS, a representative sample of 33 local clusters ($z < 0.2$) drawn from the REFLEX catalogue and observed with \textit{XMM-Newton} satellite, and three large samples of simulated clusters at redshift zero extracted from hydrodynamical simulations \citep{bib:borgani, bib:nagai, bib:piffaretti}. The fit on the profile was performed in the radial range [0.03-4]$\times R_{500}$. In this radial range, the observed profile is limited to $R_{500}$ and the region outside this radius was extrapolated according to the predictions from numerical simulations. They describe the resulting pressure profiles as universal, since it fits well both data from simulations and observations, with parameters listed in Table \ref{Tab:param_P_lowz}, fifth row. 

The \textit{Planck} Collaboration (\citealt{bib:planck_gnfw}, P13) compared the median gNFW pressure profile of 62 nearby ($z<0.5$) massive observed clusters with the A10 profile, finding that there is a very good agreement in the cluster intermediate radii between the two. However, within the core, i.e. $R < 0.15 R_{500}$, the observed profile lies significantly below the A10 profile. The fit was done in the radial range [0.02-3]$\times R_{500}$ and the parameters are reported in the fourth row of Table \ref{Tab:param_P_lowz}. 

\begin{figure*}
    \centering
    \includegraphics[width=.329\textwidth]{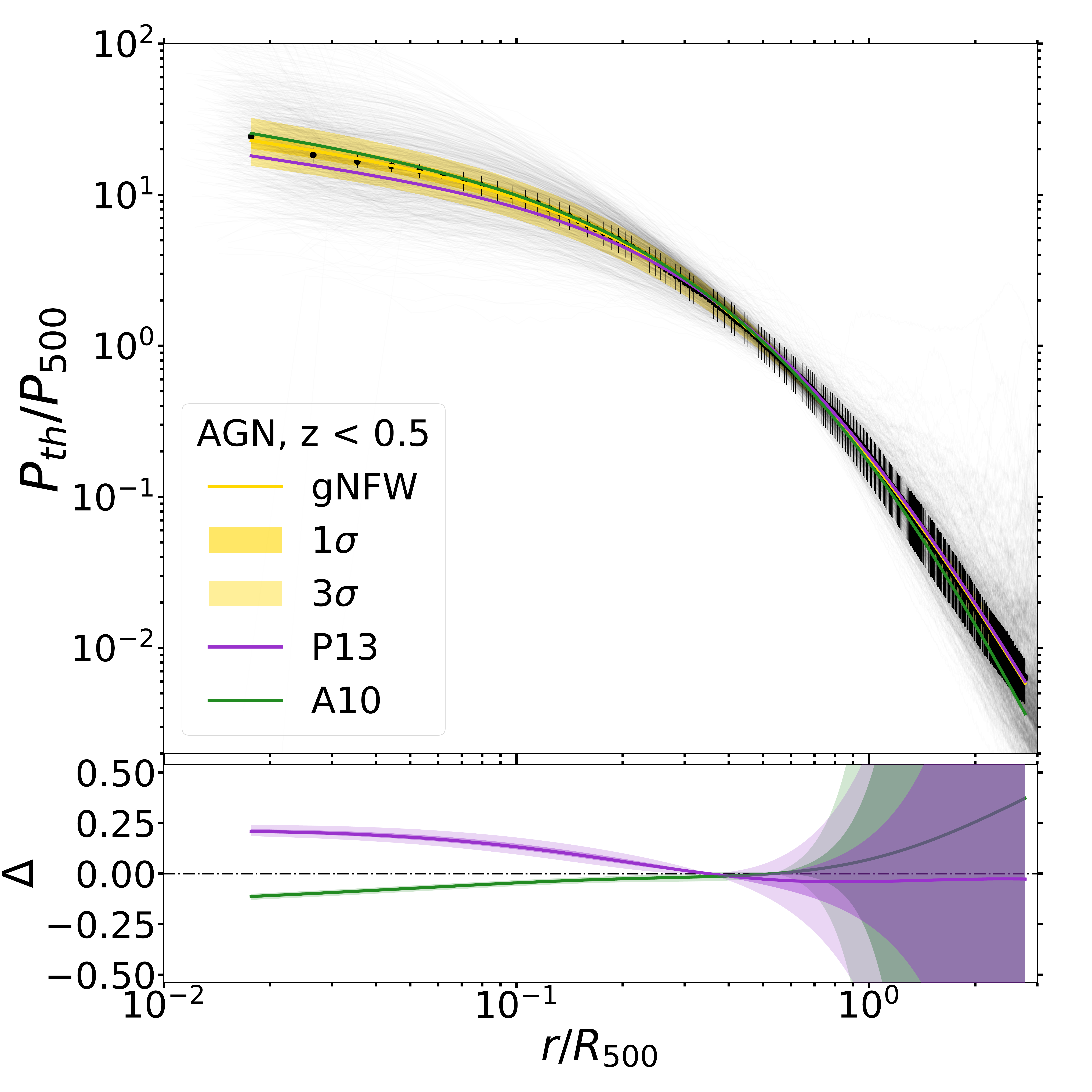}
    \includegraphics[width=.329\textwidth]{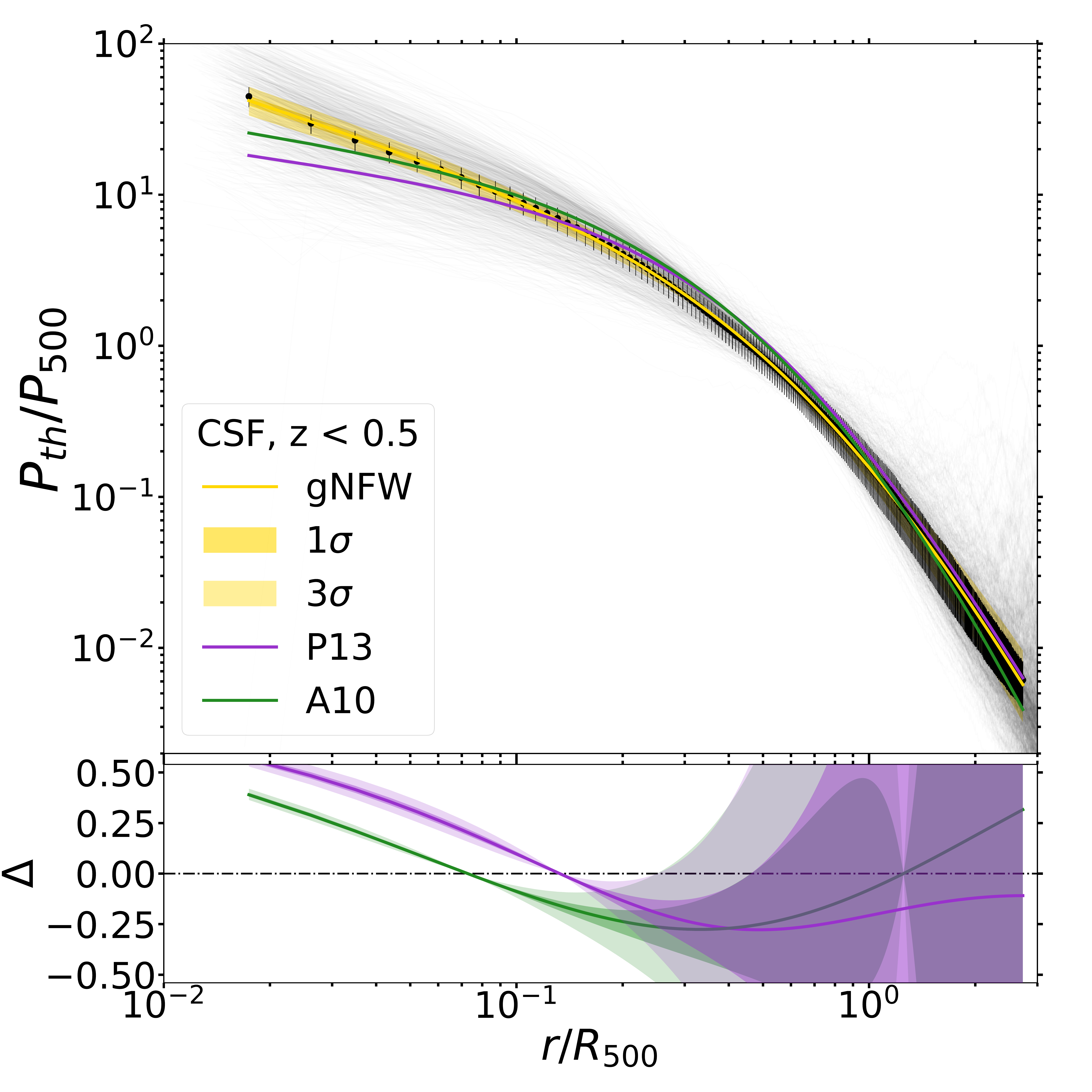} 
    \includegraphics[width=.329\textwidth]{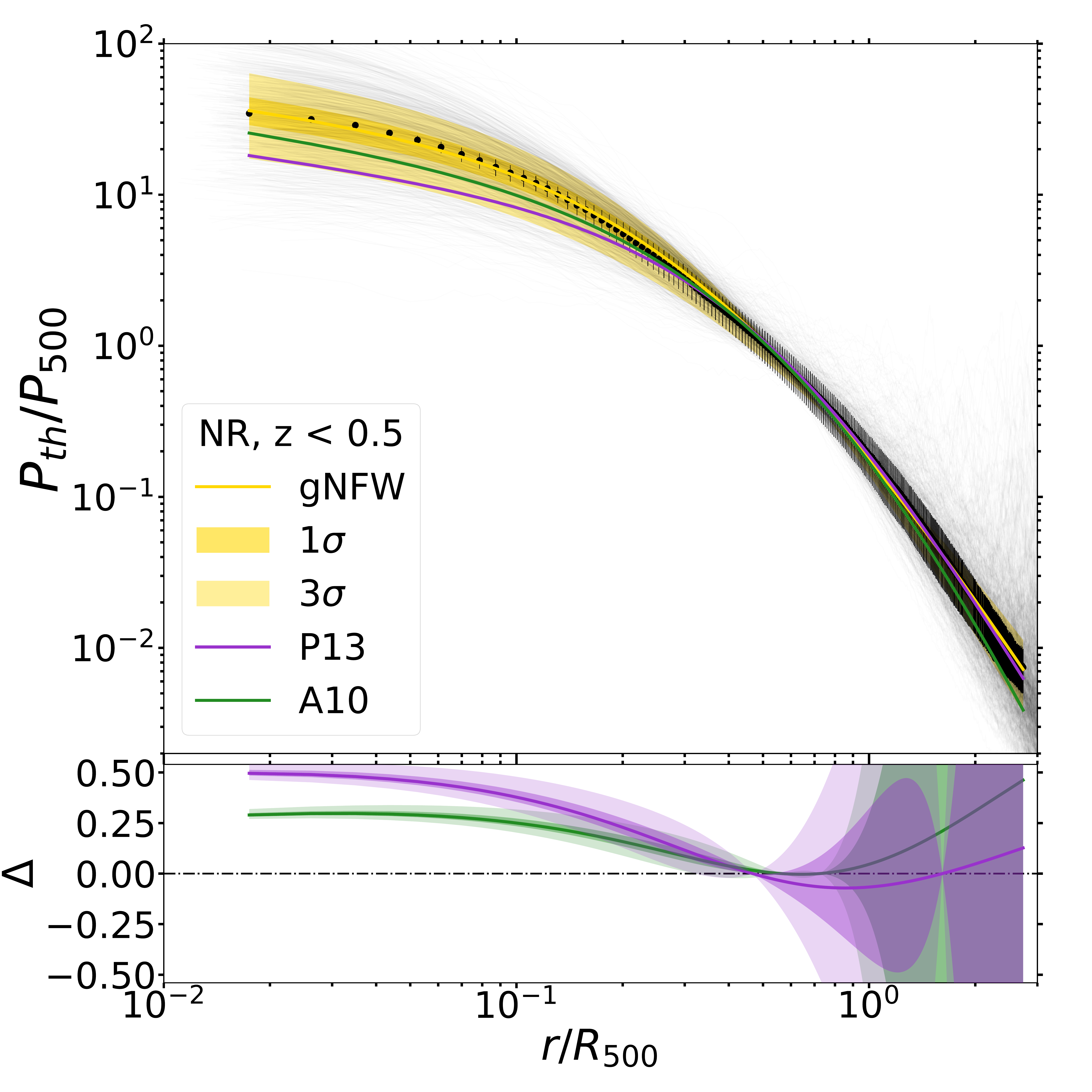} 
    \caption{The median pressure profile of the MUSIC clusters (almost 1050) at redshift z<0.5 for the AGN (left panel), CSF (central panel) and NR (right panel) simulation flavours. The grey lines represent the individual profiles for each cluster. The overall median profile is represented by black dots, with error bars computed by the MAD (Eq.(\ref{eq:mad})). The yellow area shows the variation of the best fit for the gNFW between 1 and 3 $\sigma$ intervals. The bottom panel of each plot shows the relative difference between the fit of the MUSIC profile and the corresponding profiles from A10 (green) and P13 (magenta) respectively.}
    \label{fig:planck_zlow05}
\end{figure*}

\begin{figure}
    \centering
    \includegraphics[width=.4\textwidth]{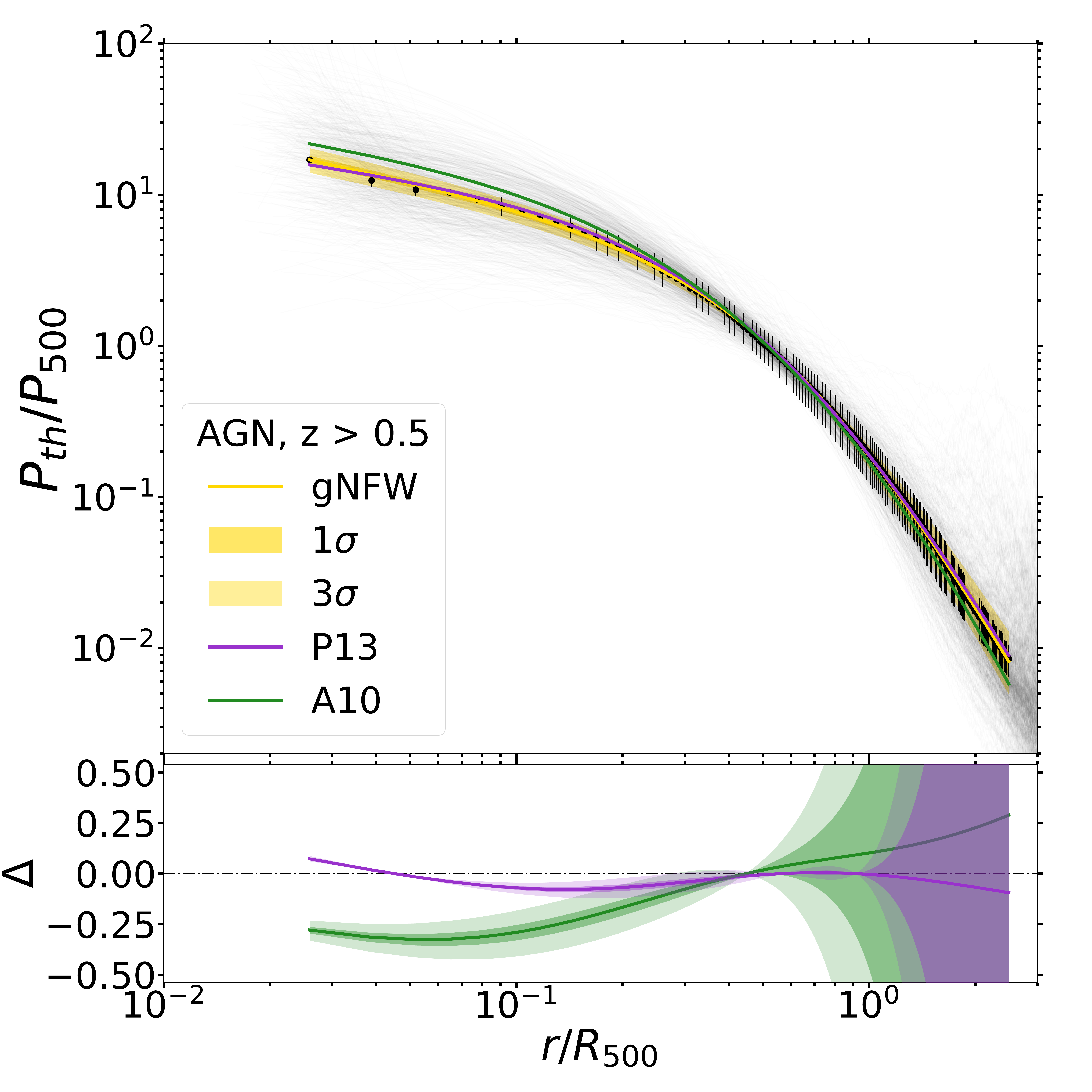}
    \caption{Same as Fig. \ref{fig:planck_zlow05} but for the MUSIC clusters at high redshifts ($z>0.5$, almost 800 clusters), for the AGN flavour only.}
\label{fig:planck_zhigh05}
\end{figure}

To compare our simulated sample with observations, we compare the median pressure profiles (dots in Fig. \ref{fig:planck_zlow05}) for all the MUSIC cluster sets at low redshifts ($z<0.5$, a sample of almost 1050 clusters), with the A10 and P13 gNFW profiles, represented in green and violet respectively. The three panels show the MUSIC median profiles for the simulation flavours, AGN, CSF and NR from left to right, with the MAD, see Eq.(\ref{eq:mad}), as associated uncertainty. The yellow line and the shaded regions represent the MUSIC pressure profile fit, with the gNFW model parameters for AGN, CSF and NR listed in the first three rows of Table \ref{Tab:param_P_lowz}. In the bottom plot of each panel we present the relative difference between the MUSIC pressure profile fit, $f_{M}$, and the profile from A10 or P13
\begin{equation}
\Delta = \frac{f_{M}-f_{A10/P13}}{f_{M}}.
\label{fig:delta}
\end{equation}
The AGN flavour is the dataset showing better agreement with both A10 and P13 profiles, especially for the A10's universal profile, a very good approximation until $R_{500}$. In the case of the CSF flavour, the MUSIC profiles are steeper starting from 0.1$R_{500}$, while in the NR case the MUSIC profiles are higher than the observed profiles within even a larger region (approximately 0.3$R_{500}$). Since only the AGN set provides a reliable description of the observed profiles, we used this set to check the redshift dependence of the universal profile, extended to the high redshift regime ($0.54<z<0.82$). As we can see from Fig. \ref{fig:planck_zhigh05}, our data match well enough both the P13 profile and the A10 one with a relative difference of the order of $0.25$. This difference is basically at the cluster core, inside $0.1 R_{500}$, and in the outskirts, beyond $2R_{500}$. Contrary to the low $z$ case, now the P13 parameters are in better agreement with our simulations. The best fit parameters for the high-$z$ case are listed in Table \ref{Tab:param_P_highz}.

\begin{figure}
    \centering
    \includegraphics[width=.4\textwidth]{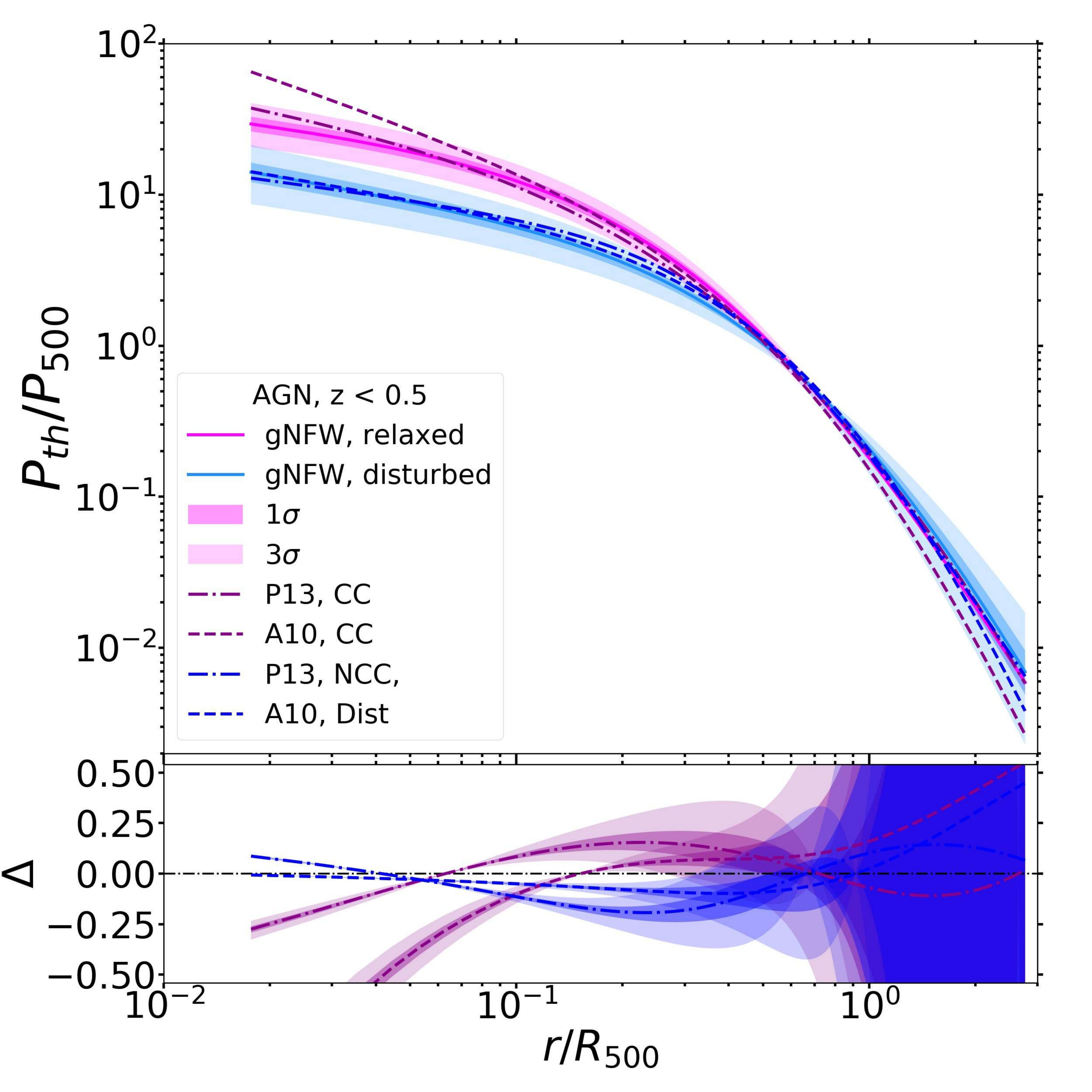}
    \caption{The best fit gNFW model for the median pressure profile of relaxed (in magenta) and disturbed (in blue) MUSIC clusters at low redshifts. The P13 profiles for cool-core and non cool-core clusters are also shown as dot-dashed lines (dark magenta and dark blue) and the A10 profiles as dashed lines. The bottom panel represents the relative difference, $\Delta$ (Eq. (\ref{fig:delta})), for relaxed and disturbed clusters only.}
\label{fig:planck_rel_dis}
\end{figure}

We have also studied whether the dynamical state of the clusters could have an impact in the comparisons with observations.
While we segregated extremely relaxed and disturbed clusters in MUSIC from the dynamical state parameters (see \ref{subsec:dyn_state}), it is not that straightforward with P13 observed data. The
dynamical state classification of P13 is based
simply on the presence or absence of a cool core \citep{PlanckCollaboration2011_b}, and assumes that a cool core represents a relaxed system (see e.g. \citet{bib:cc_cl}, although this assumption does not always hold, see \citealt{bib:biffi}). 
In A10,  the cluster dynamical state classification is somewhat  different: the classification was established in \citet{Pratt_2009}, where the subsamples were defined as cool core (10 systems), morphologically disturbed (12 systems), and neither cool core nor morphologically disturbed (11 systems). In Fig. \ref{fig:planck_rel_dis} the median pressure profiles for relaxed and disturbed clusters are plotted both at low redshifts for the AGN flavour. The parameters of the fit are listed in Table \ref{Tab:param_P_lowz}, only for the AGN flavour because is the one that better matches real observed clusters, as shown in Fig. \ref{fig:planck_zlow05}. We can see that the disturbed profile from A10 and the NCC P13 profile match very well the MUSIC disturbed profile inside $R_{500}$. This means that the disturbed clusters from A10 and from the MUSIC simulation have similar features as the NCC clusters in P13. This does not happen for the relaxed MUSIC clusters, showing a shallower slope of the profile in the cluster core with respect to the CC clusters from A10 and P13. The reason for this behaviour could be that selecting dynamically relaxed clusters based on the indicators described in Section \ref{subsec:dyn_state} does not allow us to discriminate cool-core clusters and clusters with a disturbed core. Therefore, we expect the inner slope of the MUSIC pressure profile estimated on relaxed clusters to be shallower than the one obtained by studying only CC clusters. Another possible explanation is that our AGN feedback model is too effective and expels more gas from the cluster core.

\begin{table*}
\centering
\caption{ The best fit parameters and their errors for the gNFW overall median pressure profile at low redshifts ($z<0.5$). In the first part of the table the parameters are listed for the three simulation flavours (AGN, CSF, NR), followed by the parameters from the A10 universal profile and the P13. In the second part of the table we show the best fit parameters for relaxed MUSIC clusters (AGN flavour) and for the Cool Core clusters from the A10 and P13. The description is exactly the same for the third part of the table, but in this case for the disturbed clusters and the non cool core clusters from P13.}

\begin{tabular}[t]{cccccc}
\hline
 & $P_0$   &  $c_{500}$  &  $a$  &  $b$  &  $c$  \\
\hline
 AGN & $8.87\pm0.52$ & $1.77\pm0.04$ & $1.15\pm0.03$ & $4.29\pm0.06$ & $0.29\pm0.02$ \\ 
 CSF & $2.55\pm0.10$ & $1.20\pm0.05$ & $1.14\pm0.03$ & $4.50\pm0.09$ & $0.74\pm0.02$ \\ 
 NR  & $20.00\pm0.20$ & $2.67\pm0.13$ & $1.07\pm0.05$ & $3.40\pm0.09$ & $0.23\pm0.03$ \\ 
 P13 & 6.41 & 1.81 & 1.33 & 4.13 & 0.31 \\ 
 A10 & $8.403$ & $1.177$ & $1.051$ & $5.491$ & $0.308$ \\ 
\hline
 AGN Relaxed & $13.40\pm0.72$ & $2.39\pm0.03$ & $1.24\pm0.03$ & $3.95\pm0.04$ &  $0.27\pm0.02$\\ 
 P13 CC & 11.82 & 0.60 & 0.76 & 6.58 & 0.31 \\ 
 A10 CC & 3.25 & 1.13 & 1.22 & 5.49 & 0.77 \\ 
\hline
 AGN Disturbed & $3.29\pm0.24$ & $1.35\pm0.09$ & $1.38\pm0.07$ & $4.35\pm0.16$ & $0.39\pm0.03$] \\ 
 P13 NCC & 4.72 & 2.19 & 1.82 & 3.62 & 0.31 \\ 
 A10 Dist. & 3.20 & 1.08 & 1.41 & 5.49 & 0.38\\ 
\hline
\end{tabular} 
\label{Tab:param_P_lowz}
\end{table*}

\begin{table*}
\centering
\caption{Same as Table \ref{Tab:param_P_lowz} but for all the MUSIC clusters at high redshifts ($z>0.5$) from AGN simulations (first row). The last two rows show the parameters when considering only relaxed and only disturbed clusters.}

\begin{tabular}[t]{cccccc}
\hline
  & $P_0$   &  $c_{500}$  &  $a$  &  $b$  &  $c$  \\ 
\hline
 AGN & $3.57\pm0.13$ & $1.62\pm0.05$ & $1.54\pm0.05$ & $4.18\pm0.07$ & $0.50\pm0.01$ \\ 
\hline
 AGN Relaxed & $4.52\pm0.21$ & $1.88\pm0.05$ & $1.61\pm0.06$ & $4.00\pm0.07$ & $0.53\pm0.02$ \\ 
\hline
 AGN Disturbed & $1.27\pm0.17$ & $1.44\pm0.06$ & $3.67\pm0.74$ & $3.93\pm0.13$ & $0.67\pm0.05$ \\ 
\hline
\end{tabular} 
\label{Tab:param_P_highz}
\end{table*}

\subsection{Electron density profile}
\label{sec:ne}

The 3D electron density profiles, $n_{\rm e}(r)$, of our simulated clusters are modeled using the analytical function proposed by \citet{bib:vikh}
\begin{equation}
n_{\rm p} n_{\rm e}(r) = n_0^2 \frac{(r/r_{\rm c})^{-a}}{ (1 + (r/r_{\rm c})^2)^{3b-a/2}} \frac{1}{ (1 + (r/r_{\rm s})^c)^{e/c} } + 
\label{eq:ne_vikh}
\end{equation}
\begin{equation*}
  + \frac{n_{02}^2}{ (1 + (r/r_{\rm c2})^{2})^{3b_2} },
\end{equation*}
which is a modification of the $\beta-$model \citep{bib:cavaliere} to represent the observed features of X-ray observations. It is based on two terms. The first term represents a cuspy profile near the cluster centre plus another power law to describe the steepening of the profile for $r> r_s$, the extra parameter $c$ controls the width of the transition region. The second term is another $\beta$-model with a small central core to make the function flexible to fit the data in the central region of the clusters. All clusters profiles were fitted using the same parameter constraints as in \citet{bib:vikh}, that is fixing $c = 3$ and $e < 5$.

\subsection{Temperature profile}
\label{sec:T}

The temperature profile is estimated as the mass-weighted average over an ensemble of gas particles within each spherical shell:
\begin{equation}
T = \frac{\sum_i (m_i \times T_i)}{\sum_i m_i}.
\end{equation}
where $i$ run over each particle inside a spherical bin. We considered only particles with temperature $kT > 0.5$ keV to account only for the X-ray emitting gas. 
The analytical model describing the mass-weighted temperature was introduced also by \citet{bib:vikh}
\begin{equation}
T(r) = T_0 \times \frac{x + \tau}{x+1} \times \frac{(r/r_{\rm t})^{-a}}{ (1 + (r/r_{\rm t})^b)^{c/b} }
\label{eq:T_vikh}
\end{equation}
\begin{equation*}
x = (r/r_{\rm cool})^{a_{\rm cool}}.
\end{equation*}
The radial temperature profile has a broad peak at $r < 0.1 R_{500}$ and decreases at larger radii, there is also a temperature decline towards the cluster center, probably because of the presence of radiative cooling, represented by the central expression. Outside the central cooling region, the temperature profile is represented as a broken power law with a transition region, the last term, where $r_{\rm t}$ is a scale radius. This model has 8 free parameters, none of them was fixed. 

The mass weighted temperature is the value that better relates to the mass of the cluster, in fact it reflects the gravitational potential of the system \citep{Biffi2014}. Nevertheless, there are various other ways of estimating the cluster temperature, for instance weighting the temperature by the X-ray emission of each gas particle, such as the spectroscopic-like temperature $T_{\rm sl}$ \citep{mazzotta2004}, in order to better explore the X-ray observable properties of simulated galaxy clusters and to compare against real observations \citep{rasia14}. \citet{Biffi2014} have computed the spectroscopic-like temperature for the MUSIC clusters by fitting the emission spectra from SPH particles in the energy band $0.5-10 $ keV. They found that this temperature is, on average, lower than the mass weighted temperature.

\section{Results for the full MUSIC sample}
\label{sec:res}

In this section we present the results for the hydrostatic masses defined by Eq.(\ref{eq:Mhe_P}) and (\ref{eq:Mhe_T}), for the complete MUSIC sample. We will devote Section \ref{sec:nika2} and Appendix \ref{app:nika2} to present  the specific results for the NIKA2 twin sample. As we mentioned above, the HE mass estimates, and consequently, the mass bias, depend on different factors: the observable quantites used, the simulation flavour, the numerical method used to estimate the spatial gradients, and finally the redshift and the dynamical state of the clusters. In this section we explore how the HE masses and biases depend on these factors.

\subsection{Methods to estimate the HE masses}
\label{subsec:methods}

In this paper, we use two methods to estimate the HE masses accordingly to the ways of computing the derivatives in Eq.(\ref{eq:Mhe_P}) and Eq.(\ref{eq:Mhe_T}). In the first approach, the ICM radial profiles are estimated from the SPH gas particles within each spherical bin. Then the derivatives can be directly computed numerically from the binned profiles. This is a more direct estimation of the gradients but can also suffer from noise associated to the bin size and particle sampling. An alternative estimation is to first fit the numerical profiles by the analytical functions (described in Section \ref{sec:profiles}) and take the derivative from the fitted function. 

The first approach was applied in several studies \citep{bib:lau, bib:ameglio, bib:semb, bib:battaglia, bib:nelson, bib:biffi, bib:shi, bib:martizzi, bib:lebrun, bib:cialone}. In this work, each cluster is divided in spherical shells with 100 kpc width, which has been found to be the optimal binning for our purpose. This procedure corresponds to smoothing the profile, still keeping its most important features when comparing with the original binning.

The second method is based on the analytical derivative of the fitting functions of the ICM radial profiles. Also in this case, several other works have made use of this method \citep{bib:piffaretti, bib:meneghetti, bib:kay, bib:rasia, bib:mccarthy, bib:gupta, bib:ruppin, bib:pearce, bib:rasia19, barnes2020}. The radial profiles of each cluster thermodynamic quantity are fitted with parametric models (see Sections \ref{subsec:Pprof}, \ref{sec:ne}, \ref{sec:T}) using the Python function \textit{curve\_fit}. The profile bootstrap errors have been estimated instead of the MADs, which mainly in the outskirts show too large values, often of the order of the median quantity. The fits are done in the radial range $[0.1-1] \times R_{500}$. So, we neglect the core and the outskirts of the clusters, since in observed cores there is a large variation from cluster to cluster and it is still challenging to reach the external regions in X-ray. We fix a maximum value of 10 for the reduced $\chi^2$, hereafter $\hat{\chi}^2$, to include those fits which are still, visually, a good approximation to the profiles due to the slight variations and the small errors. In this way, we have almost 50\% of the total number of clusters which have reliable fits ($\hat{\chi}^2 \leq 10$). The group of reliable fit clusters changes depending on the HE mass chosen (X-ray or SZ). In fact, having a cluster a reliable fit for one of the thermodynamic quantities does not necessary means that it has a good fit also for the other quantities. 

\subsection{Hydrostatic mass estimates}
\label{sec:results_mass}

\begin{figure}
    \includegraphics[width=.45\textwidth]{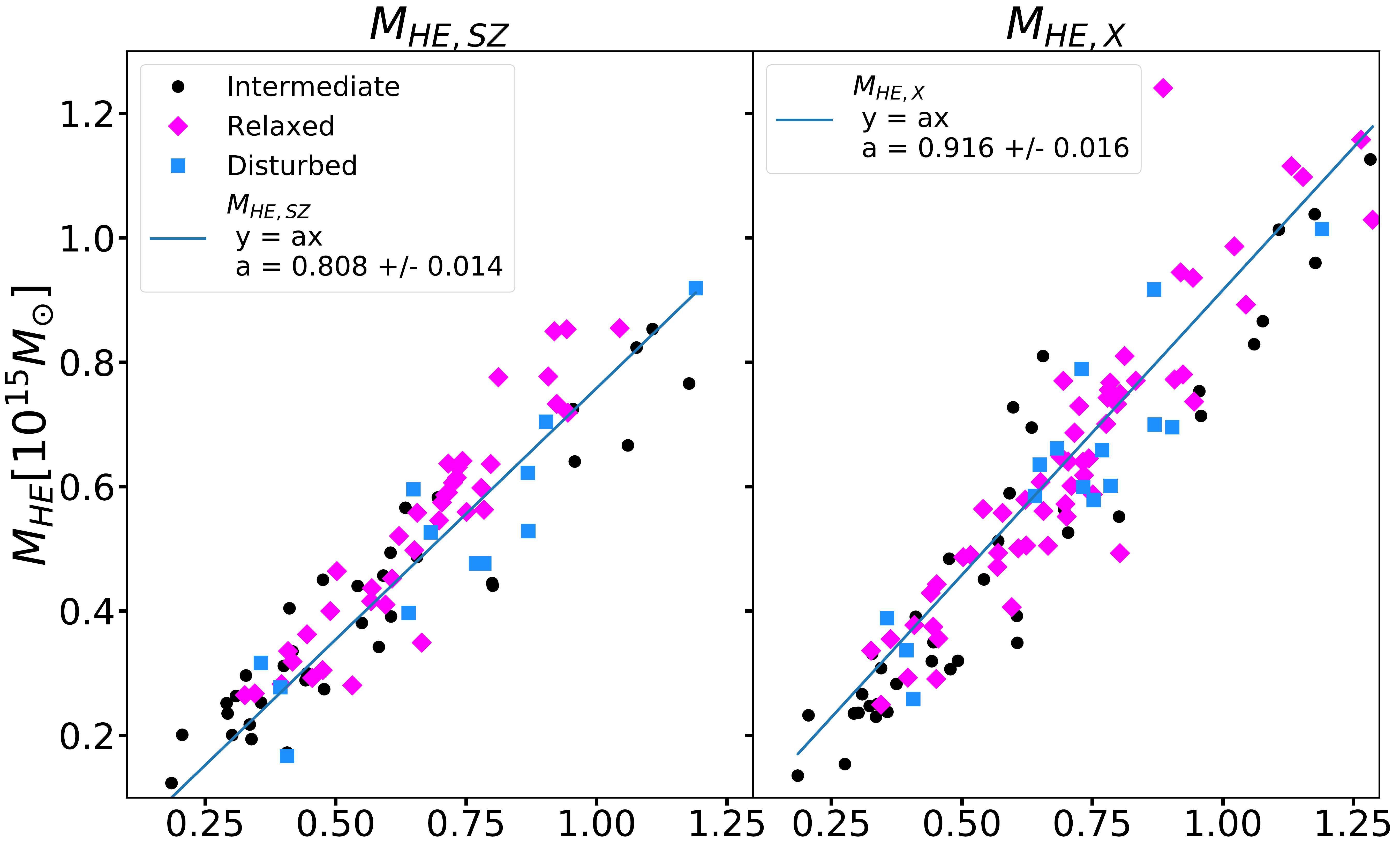}
    \includegraphics[width=.45\textwidth]{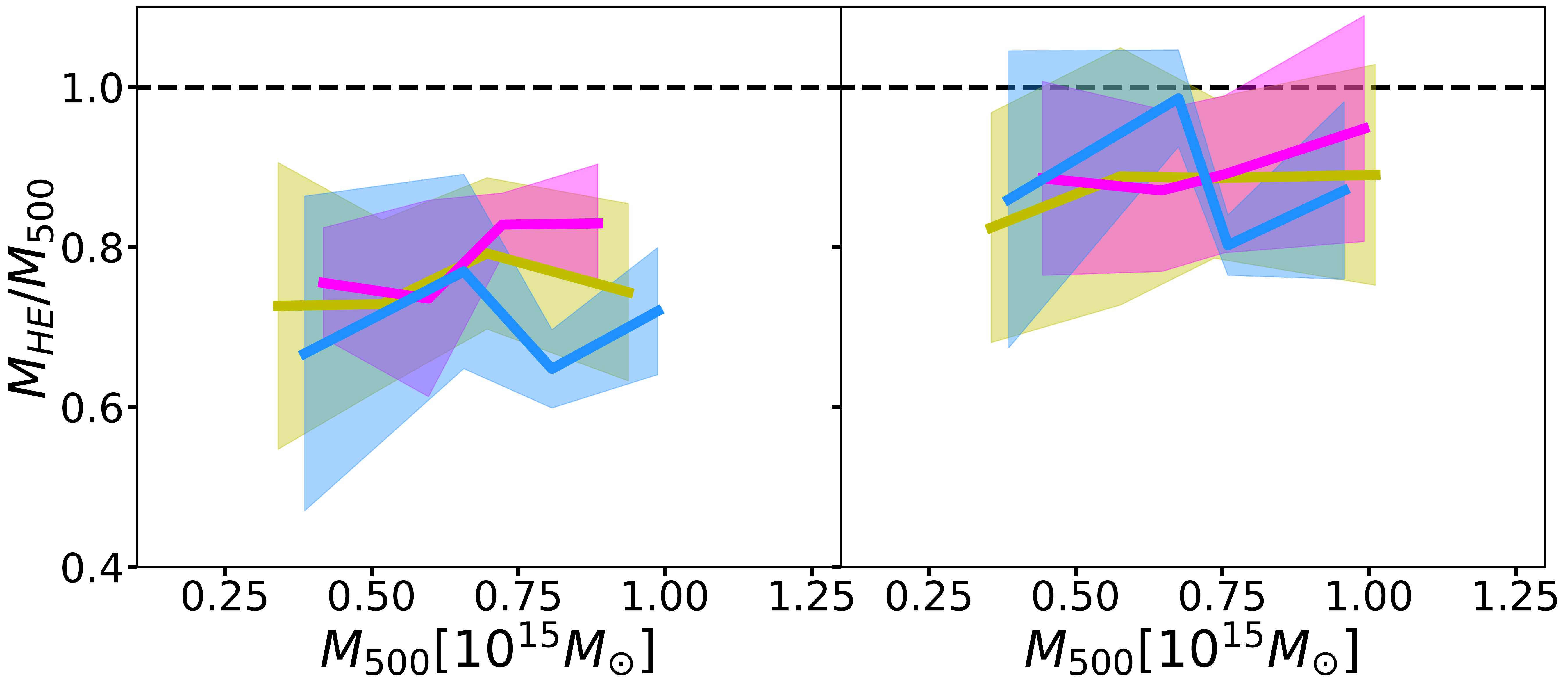}
 	\caption{First row. The HE masses $M_{\rm HE, SZ}$ and $M_{\rm HE, X}$ as a function of $M_{500}$ for the AGN simulations at $z = 0$ using the analytical fitting method. Only clusters with reliable fits are shown. A linear fit is performed and the value of the slopes are listed in the plots. The relaxed clusters are represented as magenta diamonds, the disturbed ones as blue squares and the intermediate as black circles. Second row. A binned average of the ratio of $M_{\rm HE}$ (the same of the first row) and $M_{500}$ is represented as a function of $M_{500}$. The magenta and blue lines are the averages of the relaxed and disturbed clusters respectively, while the yellow one the average over all the clusters. The shaded regions show the 1$\sigma$ intervals.}
    \label{fig:an_deriv_mhe_m500}
\end{figure}

In the first row of Fig. \ref{fig:an_deriv_mhe_m500} we show the HE masses computed at $R_{500}$, from SZ and X-ray observables estimated using the analytical fitting method, only for clusters that present reliable fits ($\hat{\chi}^2 < 10$) and for the AGN simulation flavour. They are plotted as a function of the cluster true mass $M_{500}$ at $z = 0$. The HE mass is proportional to the true mass. Clusters at other redshifts show a similar behaviour, as \citet{bib:lebrun} also find. We fit the linear relation between the HE mass and the true mass. The same analysis was done for all the simulation flavours and redshift bins.
The values of the slope $a$, which is equivalent to the bias, $1-b$, for all the simulation flavours and redshifts can be seen in Table \ref{Tab:an_deriv_slopes_good}. We do not find any clear dependence on redshift, with values of $1-b$ that are around $0.8-0.9$ for all the flavours, roughly in agreement also with the results using the numerical derivative method, not shown here. Taking into account only clusters with reliable fits, leads to smaller slope values with respect to the numerical derivative method and to the situation in which we do not neglect any clusters with bad fits. In particular, in the last two methods usually the slope errors are larger than the only reliable fit method, and ultimately the results are compatible within $1\sigma$. However, fitting the profiles implies that we are not sensitive to extreme local ICM fluctuations, meaning that we are neglecting information about the clusters and that we could underestimate the HE mass. On the other hand, gas substructures and fluctuations do influence the bias, possibly leading to an overestimation of it.

\begin{table*}
    \caption{Slopes and $1\sigma$ errors from the fit $M_{\rm HE} = aM_{500} $ in Fig. \ref{fig:an_deriv_mhe_m500} at all the redshifts and for all the simulation flavour with the analytical fitting method using only the clusters with reliable fits.}
    \begin{tabular}{ccccccc} 
 \textbf{$z$} & \multicolumn{3}{c}{\textbf{Slope $a_{\rm SZ}$}} & \multicolumn{3}{c}{\textbf{Slope $a_{\rm X}$}}\\
      &  AGN   &  CSF   &  NR  &  AGN   &  CSF   &  NR  \\
\hline
0.0  & $0.808 \pm 0.014$ & $0.790 \pm 0.013$ & $0.763 \pm 0.016$ &  $0.916 \pm 0.016$  & $0.909 \pm 0.015$ & $0.890 \pm 0.018$\\
0.11 & $0.819 \pm 0.014$ & $0.800 \pm 0.014$ & $0.767 \pm 0.016$ &  $0.914 \pm 0.012$  & $0.922 \pm 0.015$ & $0.897 \pm 0.014$\\
0.33 & $0.788 \pm 0.015$ & $0.787 \pm 0.013$ & $0.731 \pm 0.010$ &  $0.876 \pm 0.015$  & $0.860 \pm 0.013$ & $0.871 \pm 0.018$\\
0.43 & $0.783 \pm 0.016$ & $0.766 \pm 0.010$ & $0.716 \pm 0.011$ &  $0.907 \pm 0.017$  & $0.886 \pm 0.013$ & $0.889 \pm 0.018$\\
0.54 & $0.758 \pm 0.014$ & $0.739 \pm 0.012$ & $0.758 \pm 0.015$ &  $0.880 \pm 0.015$  & $0.886 \pm 0.011$ & $0.855 \pm 0.014$\\ 
0.67 & $0.778 \pm 0.013$ & $0.752 \pm 0.009$ & $0.764 \pm 0.016$ &  $0.898 \pm 0.017$  & $0.923 \pm 0.013$ & $0.983 \pm 0.026$\\  
0.82 & $0.763 \pm 0.014$ & $0.789 \pm 0.014$ & $0.712 \pm 0.010$ &  $0.902 \pm 0.018$  & $0.895 \pm 0.015$ & $0.902 \pm 0.021$\\ 
\hline
\end{tabular}
\label{Tab:an_deriv_slopes_good}
\end{table*}

\subsection{Hydrostatic mass biases}
\label{sec:results_bias}

From the HE mass, the bias is estimated using Eq. (\ref{eq:bias}). According to this definition, a mass bias value of 0 suggests a HE mass equal to the true mass, therefore the hydrostatic equilibrium represents a good cluster mass approximation. This approximation does not account for contributions like, for instance, the non thermal pressure (see Section \ref{subsec:P_nth}).

At redshift 0, using the analytical fitting and the only reliable fits, we find, at $R_{500}$, $b_{\rm SZ}=0.23^{+0.14}_{-0.09}$ and $b_{\rm X}=0.14^{+0.11}_{-0.13}$ (represented in Fig. \ref{fig:sims_b}) for the AGN flavour. These results are given as median and 16th and 84th percentiles, since the biases distributions are not Gaussian, see Appendix \ref{app:distributions} for more details. The biases at different redshifts and for different simulation flavour are listed in Tab. \ref{tab:res_bias}. We can see that the X-ray hydrostatic mass tends to give a better (closer) estimation of the real mass, with respect to the SZ one. This is mainly due to the way of computing the profiles of the gas, see Section \ref{subsec:ideal_gas_law}, but the two are consistent within 1 $\sigma$.

\subsubsection{Mass bias dependence on dynamical state}
\label{sec:morph_param_num}

\begin{figure}
	\includegraphics[width=.47\textwidth]{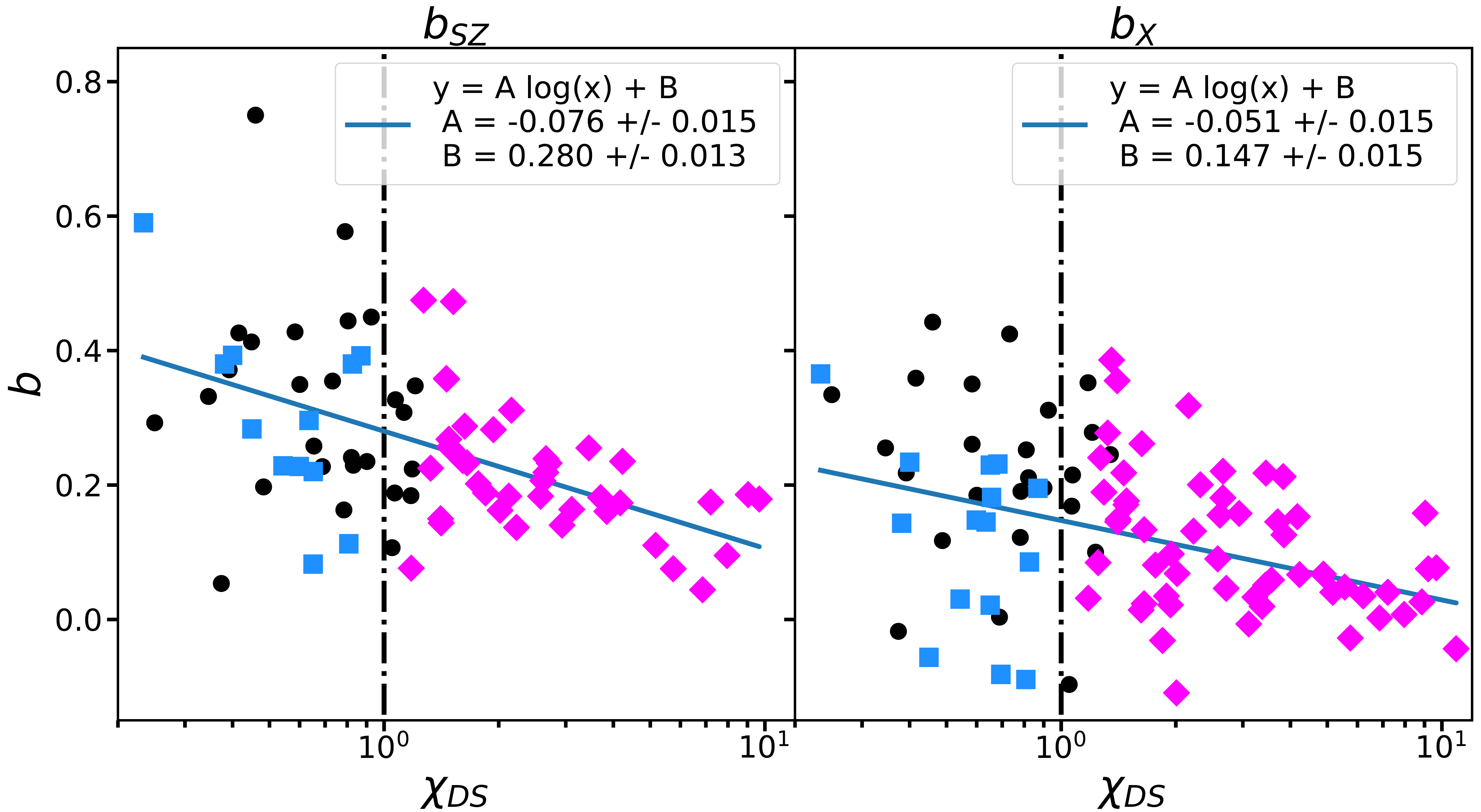}
  \caption{ The dependence of the SZ bias (left) and the X-ray bias (right) on the relaxation parameter $\chi_{\rm DS}$. The vertical dot-dashed line represents the threshold dividing the dynamically relaxed cluster (right) from disturbed ones (left). In the legend, the best fit (blue solid line) values and their errors are shown. Following the classification of the 3D morphological parameters, Section \ref{subsec:dyn_state}, the relaxed clusters are represented as magenta diamonds, the blue squares are the disturbed ones, and the intermediate corresponds to black circles.}
 \label{fig:b_vs_chi}
\end{figure}

In order to study the link between the HE mass bias and the cluster dynamical state, we test the dependence on the relaxation parameter, $\chi_{\rm DS}$ (see definition in \ref{subsec:dyn_state}). All of the clusters that have a $\chi_{\rm DS}>1$ are dynamically relaxed. The biases as a function of $\chi_{\rm DS}$, at redshift 0, are shown in Fig. \ref{fig:b_vs_chi} for the analytical fitting method, only for reliable fits. The AGN flavour is shown here, but the situation is similar for CSF and NR. The relaxed clusters present a lower scatter, compared to the other cases and also a smaller overall HE bias, as it is shown in this plot and in Table \ref{tab:res_bias}. For the SZ and X-ray bias the Pearson correlation coefficient is respectively -0.4 and -0.3, resulting in a moderate correlation and the slope of the linear relation between the bias and $\chi_{\rm DS}$ is smaller and different from zero at 3-sigma level. The latter observation is true also for the numerical derivative method (which employs all the clusters of the sample), yet the correlation is weaker, around -0.2 for both biases.

\subsubsection{Mass bias dependence on radial profile}

\begin{figure*}
	\includegraphics[width=.8\textwidth]{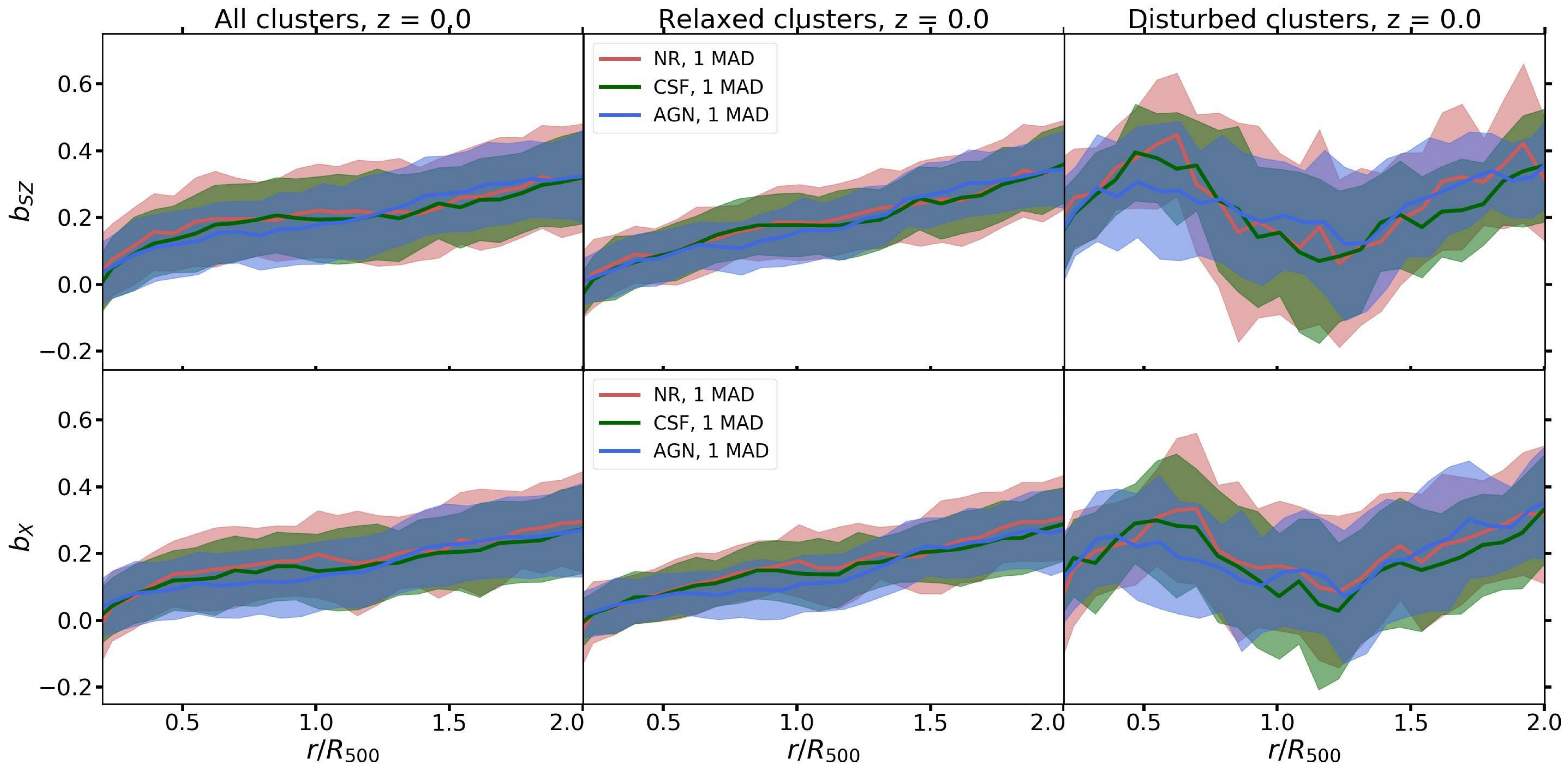} 
	  \caption{ The median SZ bias radial profile (upper row) and the corresponding one for the X-ray bias (bottom row) for $z=0$ and the 3 simulated flavours. In each panel the median profile is plotted with a blue line for AGN, the CSF profile with a green line and the NR with a red line. The shaded regions represent the 1 MAD interval for each case. In the left panels the bias is estimated over all the MUSIC clusters, in the middle, over only the relaxed ones and in the right panel only over the disturbed clusters.}
 \label{fig:radial_bias}
\end{figure*}

The numerical derivative method is powerful to explore the HE mass bias even in the large radial range. The median radial profiles and the MADs of $b_{\rm SZ}$ and $b_{\rm X}$ at $z = 0$ are shown in Fig. \ref{fig:radial_bias}, from 0.2 to 2 $R_{500}$. In the left panel of each row there is the median profile over all the clusters, in the middle only relaxed and in the right only disturbed clusters. The bias profiles from all and only relaxed classes are very regular and, as expected, increase at large radii, as found also in other works (e.g., \citealt{bib:meneghetti, bib:cialone, bib:rasia19}). This could be most likely the presence of higher non-thermal processes in the cluster outskirts, see Section \ref{subsec:P_nth}. On the contrary, the trend of the disturbed clusters has a lot of variations and a large scatter, with a dip between $R_{500}$ and $1.5R_{500}$, which seems not due to the non-thermal pressure, as a matter of fact it is still present after applying the non-thermal correction, see Section \ref{subsec:P_nth}. We will study in deep this behaviour in a future work, but we attempt here a possible explanation. It could be due to merger shocks which could boost either the ICM temperature (and therefore $M_{\rm HE}$) and consequently the gas pressure. 

\subsubsection{Bias dependence on baryon models}
\label{sec:baryon_models}

Taking advantage of exploring the same objects with different flavours, we are able to compare HE mass bias values, without worrying about the simulation features, like resolution, integrating box size, cluster mass range and number of objects, focusing only on the differences due to the physics included in the simulation. At redshift 0 and at $R_{500}$, using the analytical fitting and the only reliable fits, we find $b_{\rm SZ}=0.23^{+0.14}_{-0.09}$ and $b_{\rm X}=0.14^{+0.11}_{-0.13}$ for the AGN flavour, $b_{\rm SZ}=0.26^{+0.12}_{-0.10}$ and $b_{\rm X}=0.14^{+0.16}_{-0.11}$ for the CSF flavour and $b_{\rm SZ}=0.27^{+0.15}_{-0.11}$ and $b_{\rm X}=0.15^{+0.14}_{-0.13}$ for NR. As we can see, the simulation flavour does not influence much the bias value, as it is clear also in Table \ref{tab:res_bias}. In Fig. \ref{fig:radial_bias}, the radial profiles of both biases are represented for each flavour. Also in this case, the all clusters median profiles are not affected by the baryonic physics. 
From the middle and right panels, we see that this trend is common also to the relaxed clusters while for the disturbed objects there is a tendency for the AGN runs to have a smaller bias, although the profiles are all consistent within the scatter.

\subsubsection{The bias dependence on redshift}
\label{sec:b_dep_z}

\begin{figure}
	\includegraphics[width=.5\textwidth]{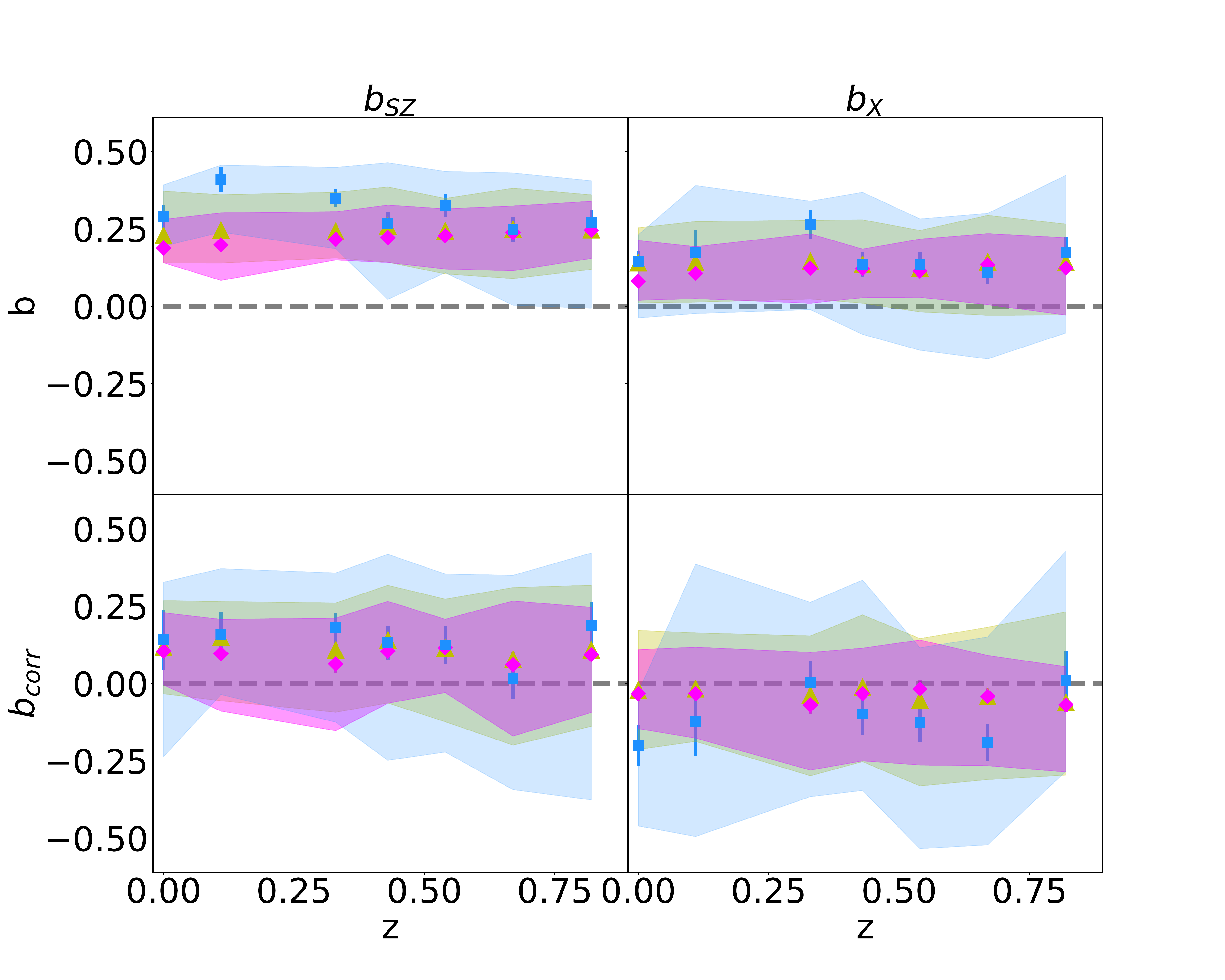}
  \caption{ The redshift evolution of the biases, $b_ {\rm SZ}$ (left panels) and $b_{\rm X}$ (right panels), from the AGN simulations. The reliable fits have been used to compute them. The median values of the bias for all clusters, relaxed and disturbed are represented with yellow triangles, magenta diamonds and blue squares respectively. The error bars correspond to bootstrap errors. The shaded regions represent the $16^{th}$ and $84^{th}$ percentiles. In the bottom panels, the corrected bias factors $b_{\rm corr}$ (see Section \ref{subsec:P_nth}) are also plotted. }
 \label{fig:b_vs_z_model_good}
\end{figure}

The bias dependence on the redshift is presented in the first row of Fig. \ref{fig:b_vs_z_model_good}. The results are from the analytical fitting method with only the reliable fits and the AGN flavour (in the case of the other flavours the situation is similar). With the current errors, we do not detect any dependence with the redshift, for either the bias (as in \citealt{ bib:lebrun, Henson2017, bib:salvati, Koukoufilippas2020}) or its uncertainties. In all cases, the disturbed clusters have the largest errors, as already pointed out in Section \ref{sec:morph_param_num}. We see the same behaviour for the numerical derivative method, with the only difference that the disturbed clusters percentiles are almost twice larger than the percentiles shown in the analytical fitting method. In this figure the bootstrap error is represented too, for the three cases.

\subsubsection{The bias mass dependence}

A binned average of the quantity $1 - b = M_{\rm HE}/M_{500}$ is represented in the second row of Fig. \ref{fig:an_deriv_mhe_m500}, as a function of $M_{500}$. We do not find any mass dependence inside the 1$\sigma$ regions, in agreement with \citealt{bib:rasia19}.

Several authors detect a dependence when using the spectroscopic-like temperature, see e.g. \citet{rasia06, bib:piffaretti, bib:lebrun, Henson2017, bib:pearce, barnes2020}. This result is mainly due to the fitting of the gas multi-temperature spectrum with a single temperature model. The plasma in the outskirts is characterized by a higher degree of substructures not yet thermalized and therefore colder and denser. This reduces the temperature evaluated at $T_{500}$. Indeed, the mass bias estimated using  the spectroscopic-like temperature, is generally higher and with a larger scatter with respect to the mass-weighted temperature case. We  have also confirmed this behaviour in our MUSIC  HE bias estimates (at redshift 0)  when using the  spectroscopic-like temperatures, mostly  because of the larger scatter of their profiles. But, again,  no  dependence of the HE bias on the cluster mass was found   when the spectroscopic-like temperature profiles are used.

\subsubsection{Correlation between $b_{\rm X}$ and $b_{\rm SZ}$}
\label{subsec:ideal_gas_law}

We find that the two bias factors are strongly correlated, with a Pearson coefficient of 0.8. To estimate the scatter between the two biases we can use the mean absolute difference 

\begin{equation}
MD = \sqrt{\frac{\sum_i (b_{\rm X,i} - b_{\rm SZ,i})^2}{N(N-1)}}
\label{eq:MD}
\end{equation}
with $N$ being  the total number of clusters. 
At all redshifts $MD$ is of the order of $10^{-1}$.

In the simulations, the pressure of each fluid particle is computed assuming the ideal gas equation of state, used also to derive $M_{\rm HE,X}$, Eq. (\ref{eq:Mhe_T}), from $M_{\rm HE,SZ}$, Eq. (\ref{eq:Mhe_P}), 
\begin{equation}
P_i = n_{\rm e,i} T_i,
\label{eq:ideal_gas}
\end{equation}
where $n_{\rm e,i}$ and $T_i$ are the electron density and the temperature of each particle of fluid. Then, the median over the spherical shell is performed, obtaining the median pressure profile. This profile, together with the median profile of the electron density and the mass-weighted temperature profile, are used in Eqs (\ref{eq:Mhe_P}) and (\ref{eq:Mhe_T}). For this reason we expect the two HE masses to be different.

To verify that the scatter between the two biases is really due to the previous consideration, we need to quantify it, for instance using
\begin{equation}
\frac{M_{\rm HE, SZ}}{M_{\rm HE, X}} = \frac{1-b_{\rm SZ}}{1 - b_{\rm X}} 
\end{equation}
which, replacing the masses with the Equations (\ref{eq:Mhe_P}) and (\ref{eq:Mhe_T}), becomes
\begin{equation}
\frac{1-b_{\rm SZ}}{1 - b_{\rm X}} = \frac{P}{n_{\rm e} T} \frac{\rm d\ln P/d\ln r}{\rm d\ln (n_e T)/d\ln r} \sim \frac{P}{n_{\rm e} T}
\label{eq:scatter_b} 
\end{equation}
where the derivatives fraction is always of the order of 1, no matter the flavour or the redshift.

We find that the profile of $P/n_{\rm e}T$ decreases very slowly from a value of almost 1 in the core. In the case of AGN, $z = 0$, the deviation from 1 of this ratio is $1 - P/n_{\rm e}T = 0.12$, but in general it is always of the order of $10^{-1}$. This is of the same order of the scatter $MD$, so we can conclude that the difference between the X-ray and SZ bias is mainly due to the use of the median and mass-weighted profiles.

\subsection{The non-thermal pressure contribution}
\label{subsec:P_nth}

The hydrostatic equilibrium approximation does not take into account non-thermal motions of the ICM, which could have a significant contribution to the pressure support of the gas within the cluster gravitational potential. The non thermal pressure contribution comes mostly from turbulence or bulk motions of the ICM, neglecting other sources such as magnetic fields or cosmic rays pressure. Recent studies have shown that it can contribute as much as 30 per cent or more to the overall gas pressure at $R_{500}$ \citep{bib:lau, bib:nelson, bib:shi, bib:biffi, bib:martizzi, angelinelli2019, bib:pearce}. This has been identified as the main origin of the cluster mass bias.

The non-thermal pressure component is modelled as 
\begin{equation}
P_{\rm nth} = \alpha(r)P_{\rm tot}
\label{eq:P_nth}
\end{equation}
assuming $P_{\rm tot} = P_{\rm th} + P_{\rm nth}$. From simulations, the non-thermal pressure \citep{bib:nelson_nth} can be estimated as 
\begin{equation}
P_{\rm nth} = \rho \sigma^2
\end{equation}
where $\sigma^2 = \sum_j \sigma_j^2$, with $ j = (x,y,z)$, is the square of the three-dimensional velocity dispersion of gas particles and
\begin{equation}
\sigma_j^2 = \frac{ \sum_i (m_i v_{i,j} - \bar{v_j})^2 }{\sum_i m_i},
\end{equation}is the velocity dispersion in each spatial direction computed from all the gas particles within each spherical shell around the cluster centre. 

Then, introducing $P_{\rm tot} $ in the formula for the HE, we can derive the corrected hydrostatic masses \citep{bib:pearce}
\begin{equation}
M_{\rm HE,SZ,corr} = \frac{1}{1 - \alpha} \left[ M_{\rm HE,SZ} - \frac{\alpha}{1 - \alpha} \frac{rP_{\rm th}}{G \mu m_p n_{\rm e}} \frac{\rm d \ln \alpha}{\rm d \ln r} \right] 
\label{eq:Msz_corr}
\end{equation}
\begin{equation}
M_{\rm HE,X,corr} = \frac{1}{1 - \alpha} \left[ M_{\rm X,SZ} - \frac{\alpha}{1 - \alpha} \frac{k_{\rm B} T r}{G \mu m_p} \frac{\rm d \ln \alpha}{\rm d \ln r} \right] .
\label{eq:Mx_corr}
\end{equation}
We note that the temperature profile used in the formula above corresponds to the mass-weighted temperature definition. For the non-thermal correction applied on HE mass using the spectroscopic-like temperature see \citet{rasia06}.

\begin{figure}
	\includegraphics[width=0.45\textwidth]{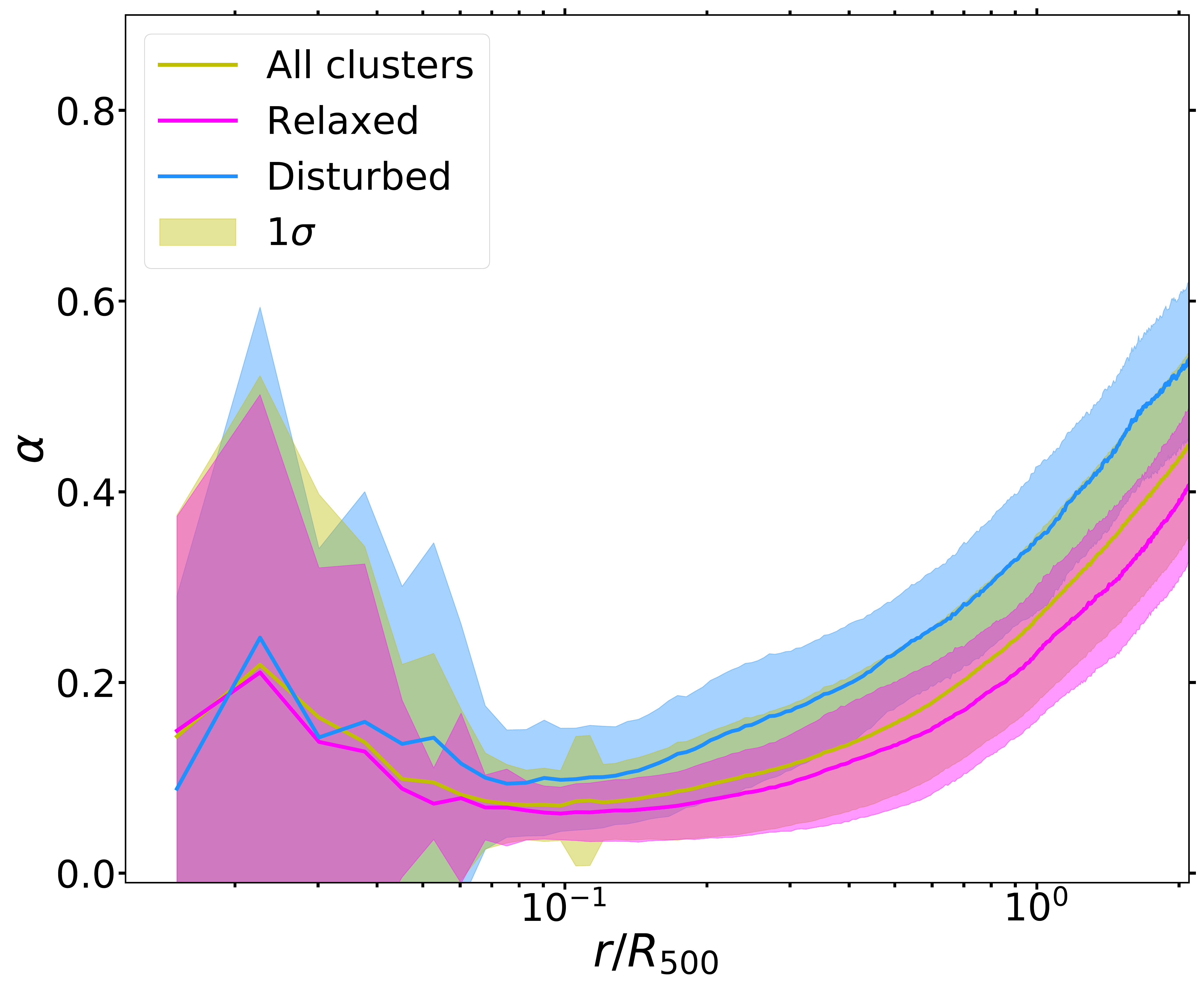} 
 \caption{The mean radial profile of $\alpha = P_{\rm nth}/P_{\rm tot}$ for MUSIC clusters in the AGN simulation at $z = 0$. The profiles for all the clusters, the only relaxed and the only disturbed are shown as yellow, magenta and blue solid lines respectively. The shaded regions represent the $1\sigma$ interval.}
 \label{fig:rad_alpha}
\end{figure}

Therefore, in order to correct the HE masses we need to know the value of the $\alpha$(r) function. We use the fitting formula originally proposed by \citet{bib:nelson_nth} for $R_{200}$, adapted to the smaller aperture radius $R_{500}$,

\begin{equation}
\alpha(r) = 1 - A \left[ 1 + \exp \left( - \left(\frac{r/R_{500}}{B} \right)^C \right) \right] 
\label{eq:alpha}
\end{equation}
where $A$, $B$ and $C$ are three free parameters \citep{bib:pearce}.

The contribution of non thermal pressure becomes significant at large radii (e.g., \citealt{bib:pearce, bib:biffi}). We confirm a similar behaviour as shown in Fig. \ref{fig:rad_alpha}, where the ratio $\alpha$ is plotted. We can see that, at $R_{500}$, it goes from $\sim 15\%$ to $\sim 40\%$, depending on the morphological state of the clusters. The most disturbed clusters have an higher contribution to $\alpha$, even if the two classes of objects (relaxed and disturbed) are consistent within their errors. Applying the non-thermal correction to the mass should result in minimising the biases. In the bottom panels of Fig. \ref{fig:b_vs_z_model_good}, indeed, the $b_{\rm X}$ and $b_{\rm SZ}$ are close to 0, but the scatter (represented by the percentiles in Table \ref{tab:res_bias}) is larger than the non corrected case due to the large scatter in the $\alpha$ profiles, in agreement with the results from \citet{bib:pearce}. In addition, we also show that there is not a significant dependence on the redshift range analyzed in this work.
Moreover, the correction on $b_{\rm X}$, which already gave an estimation of the mass nearer to 0 before the correction, is more effective than on the $b_{\rm SZ}$. This leads to zero X-ray bias for the relaxed and all clusters cases, and to an over correction for the disturbed class. Therefore, the dynamical state has an impact on the strength of the correction, but not on its radial profile, as stated before and shown in Fig. \ref{fig:rad_alpha}.

\begin{figure*}
	\includegraphics[width=.8\textwidth]{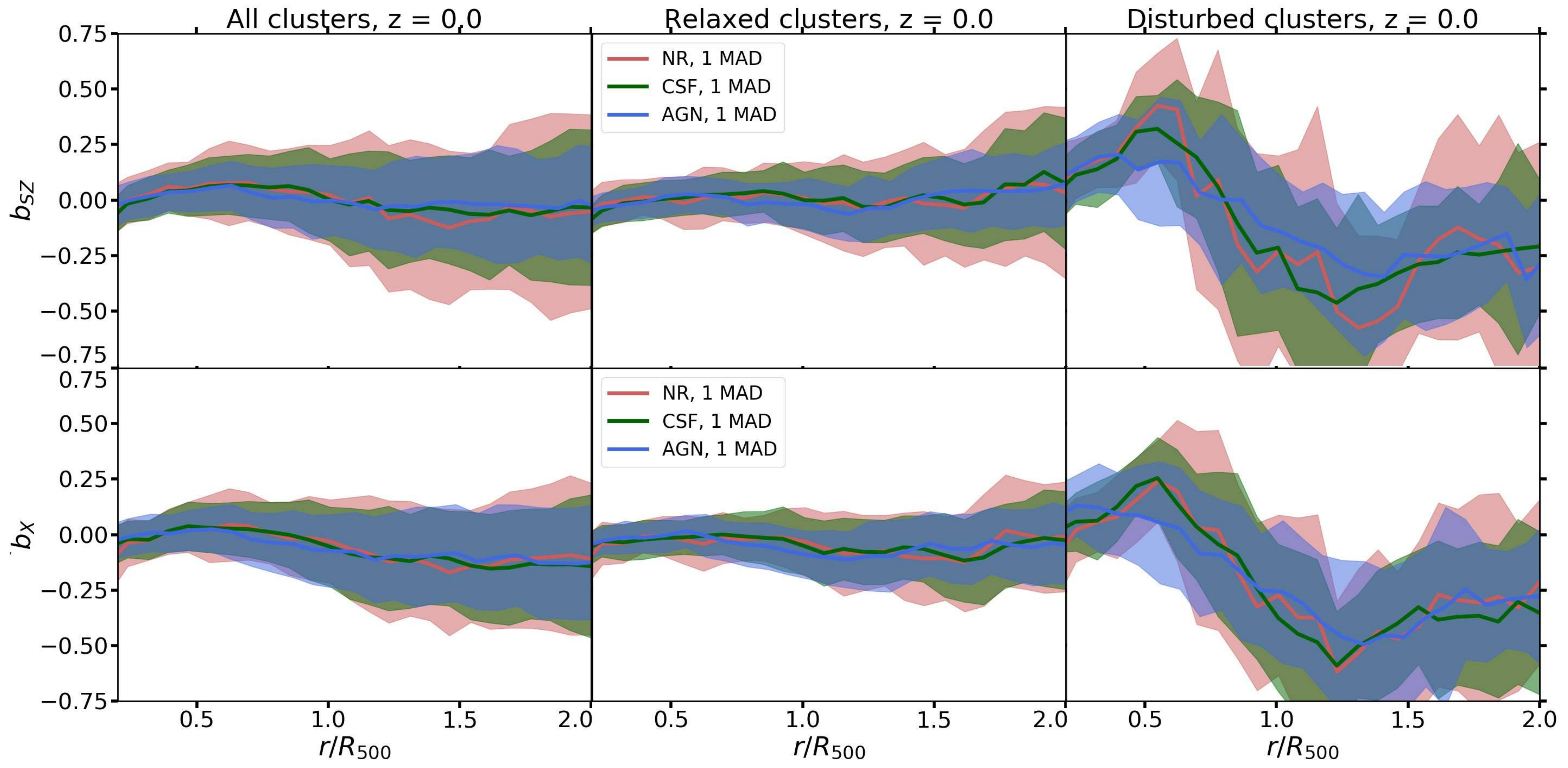}
	\caption{Same as Fig. \ref{fig:radial_bias}, but for the corrected biases.}
 \label{fig:radial_bias_corr}
\end{figure*}

The median radial profiles of the two corrected biases $b_{\rm SZ, corr}$ and $b_{\rm X, corr}$ are presented in Fig. \ref{fig:radial_bias_corr} for redshift 0 and all flavours. The increasing radial dependence of the X-ray and SZ bias was canceled when including the non thermal correction, as expected. However, there is still a large variation in the bias profiles around $R_{500}$ for the disturbed clusters, which was already present before the correction. This could be explained by a more intriguing effect, which does not depend on the non-thermal pressure, or the simple correction formula does not account for the extreme non thermal pressure from the disturbed clusters. We also note that deviations from spherical symmetry as well as the presence of substructures impacting the multi-phase structure of the gas can play a significant role in the outskirts. As we pointed out before, the scatter (represented by the MADs in this plot) is larger than the non corrected bias. Moreover, the SZ bias has larger scatter than the X-ray bias, especially for the CSF and NR flavours. The NR biases have less regular profiles with respect to the other flavours, but there is not a substantial difference between them. All the different flavours profiles are in agreement.

\section{Results from the NIKA2 twin sample}
\label{sec:nika2}

The NIKA2 twin sample \citep{bib:ruppin} is composed of 32 MUSIC clusters, see Section \ref{sec:nika2_intro}. To build the sample we chose clusters at $z=0.82$ which are not the progenitors of the clusters at $z=0.54$, as it would be also in real observations. Due to the small number of clusters, the reliable fit subsample with $\hat{\chi}^2 < 10$ having seven clusters in total, was not applied. Therefore, in the analysis, all the 32 clusters are taken into account.

The A10 and P13 pressure profiles do not quite model the data, differently from the full sample where those two profiles are acceptable even at high redshifts (Fig. \ref{fig:planck_zhigh05}, Table \ref{Tab:param_P_highz}), for the the NIKA2 median pressure profile see Fig.\ref{fig:planck_nika}.

The biases from the NIKA2 sub sample are compatible with the ones from the full sample, even though they are slightly higher, especially for the relaxed clusters (see Appendix \ref{app:nika2}). The same problem reflects on the corrected bias, their median values are shown in the second row of Fig. \ref{fig:NIKA2_an_deriv_b}.

These differences may be due to the selection strategy of the NIKA2 twin sample, in particular to the number of high mass clusters, larger in proportion than the MUSIC one (for an extended analysis see Appendix \ref{app:nika2}). Anyway, based on these analyses, we can say that the NIKA2 twin sample has compatible biases within 1 $\sigma$ with the MUSIC sample, even if it has a larger percentage number of high mass clusters.

\section{Conclusion}
\label{sec:conclusion}

The study of the bias in recovering the mass of galaxy clusters under the hydrostatic equilibrium assumption is crucial for the understanding of structures formation and cosmology. In this work the 3D ICM radial profiles of a synthetic set of almost 260 clusters from the MUSIC hydrodynamical simulations were used. The analysis has been applied on three different simulation flavours: NR, CSF and AGN. The main difference among them is the absence of radiative processes in NR, while they are included in both CSF and AGN, where also stellar and AGN feedbacks are considered, respectively. We considered seven different redshifts in the range $0.0\leq z \leq 0.82$. Moreover a MUSIC sub sample, the NIKA2 twin sample, was analysed to have some hints for the NIKA2 LPSZ.

A preliminary analysis provided the clusters classification, depending on their dynamical state. In the MUSIC dataset and in the NIKA2 twin sample we have almost 50\% of relaxed clusters. First of all, the median pressure profile of the full sample at low redshifts ($z<0.5$) was compared with two models describing observed local clusters from \textit{Planck} (P13, \citealt{bib:planck_gnfw}) and \textit{XMM-Newton} (A10, \citealt{bib:arnaud}) in order to check whether there is consistency between simulations and the analytic models used in literature.

The SZ and X-ray estimations of the HE mass were computed from the ICM thermodynamic radial profiles using two strategies to compute the gradients: the numerical and analytical fitting. The correlation between the HE masses and the true mass $M_{500}$ was studied. Therefore, the HE mass biases were estimated. Possible dependencies with redshift and dynamical state, as defined by the relaxation parameter $\chi_{\rm DS}$, were analysed, together with the bias radial profile. We also studied the impact of non-thermal pressure support arising from the bulk motion and turbulence of the gas. All these analyses were done for both the MUSIC full dataset and the NIKA2 sub sample. 

The main results of this work can be summarized as follows:

\begin{itemize}
\item In MUSIC, the AGN flavour better approximates the real clusters properties. In fact the pressure profile from the AGN simulation better matches the model from the observed clusters in P13 and A10.

\item The SZ and X-ray HE masses have a linear dependence with the true mass $M_{500}$. The slope of the fit, corresponsing to the value $(1-b)$, gives a bias $b$ of the order of 0.2, independent on the considered redshift. 

\item Also the HE biases at $R_{500}$ are of the order of 0.2, independently from X-ray or SZ formulations and derivative estimations and they do not depend on the redshift. These results are in agreement with other simulations. The full sample and the relaxed-cluster-only biases are always in agreement with each others, the disturbed population shows very high biases, with large errors. This can be simply explained because the hydrostatic equilibrium is not fully satisfied.

\item The different approaches to fit the ICM radial profiles have an impact on the scatter of the biases. The numerical derivative method gives on average a larger scatter with respect to the analytical fitting one, with the condition of excluding from the analysis the clusters with a non reliable fit.

\item The dynamical state has an impact on the scatter of the biases. Disturbed clusters show large dispersion. We find that biases depend on the continuous relaxation parameter $\chi_{\rm DS}$, which we use to describe the dynamical state of the clusters.

\item While the median radial profile of the bias for the disturbed clusters shows large fluctuations with a large scatter, the relaxed and the full sample radial biases are flatter. Interestingly, in the case of these two populations the bias profile increases radially stressing the impact of non thermal pressure contribution in the most external regions. 

\item It is evident that non thermal pressure contribution has an impact especially in the cluster outskirts. Correctly adding this contribution to the hydrostatic one, the estimated mass of the cluster really gets closer to the true one. Unfortunately, the scatter is larger, mainly due to the scatter in the fraction $P_{\rm nth}/P_{\rm tot}$, used to estimate the correction. 

\item The mass biases from the NIKA2 sub sample are compatible with the ones from the full sample, even though they are slightly higher, especially for the relaxed clusters.

\end{itemize}

The NIKA2 twin sample will help improve the analysis methodology of the ICM properties of the NIKA2 tSZ large program. Thanks to the results of this work, the sample will also help to probe different fundamental hypothesis, like spherical symmetry and hydrostatic equilibrium.

Nowadays, the main systematic uncertainty associated with cluster cosmological constraints is the uncertainty on the HE mass bias. Using the non thermal correction to obtain mass values closer to the real ones, but with a larger scatter, is not an efficient strategy, exactly as how to choose the ICM properties radial fitting approach. On the other hand, several other effects still need to be carefully considered, as temperature dishomogeneities, deviation from spherical symmetry and the presence of substructures, in order to give a more accurate measure of the HE mass bias.
 
\section*{Acknowledgements}
We thank the Referee and Stefano Andreon for their useful comments. GG, MDP and FDL acknowledge support from Sapienza Universit\'a di Roma thanks to Progetti di Ricerca Medi 2019, prot. RM11916B7540DD8D.
GY acknowledges financial support by MICIU/FEDER under project grant PGC2018-094975-C21. 
WC acknowledges supports from the European Research Council under grant number 670193 (the COSFORM project) and from the China Manned Space Program through its Space Application System. The authors wish to thank The Red Espa\~nola de Supercomputaci\'on for granting computing time in the MareNostrum Supercomputer at the BSC-CSN where the MUSIC simulations have been run.
FM, LP, FR, FK and JFMP acknowledge financial support from the French National Research Agency in the framework of the `Investissements d'avenir' program (ANR-15-IDEX-02). FR acknowledges financial supports provided by NASA through SAO Award Number SV2-82023 issued by the Chandra X-Ray Observatory Center, which is operated by the Smithsonian Astrophysical Observatory for and on behalf of NASA under contract NAS8-03060.
ER acknowledge financial contribution from ASI-INAF n. 2017-14-H.0. VB acknowledges support by the DFG project nr. 415510302.


\section*{Data Availability}

The profiles analyzed in this work were produced with the MUSIC simulations \citep{bib:semb}. These data can be accessed following the instructions on the website \url{https://music.ft.uam.es/}. The data specifically shown in this paper will be shared upon request to the authors.


\medskip

\bibliographystyle{mnras}
\bibliography{bib_tesi}


\appendix

\section{Results for the NIKA2 twin sample}
\label{app:nika2}

\begin{figure}
    \centering
    \includegraphics[width=.4\textwidth]{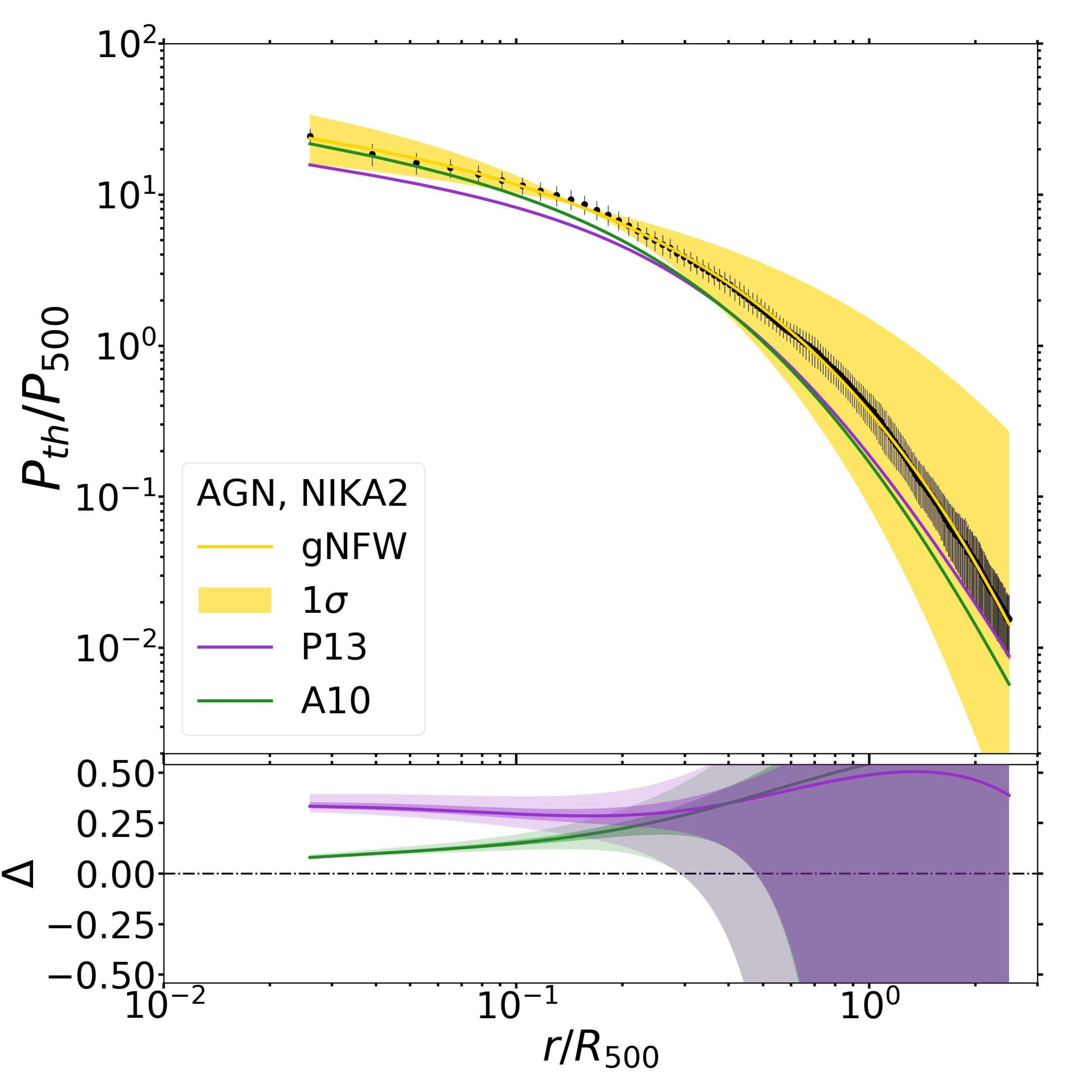}
    \caption{As Fig. \ref{fig:planck_zlow05}, the median fit to the universal pressure profile for the NIKA2 twin sample clusters (black dots). The bottom panel shows the deviation between the median profiles for NIKA2 sample and the P13 and A10 profiles.} 
\label{fig:planck_nika}
\end{figure}

The median pressure profile of the NIKA2 twin sample is plotted in Fig.\ref{fig:planck_nika} and the parameters are listed in Table \ref{Tab:param_P_nika}, first row. In this case the profiles from A10 and P13 do not approximate well the data, even though the sub sample is taken from MUSIC. To check whether this behaviour is expected, 10 sub samples were randomly extracted from MUSIC, keeping the same features in mass, redshift and dynamical state than the NIKA2 twin sample, see Section \ref{sec:sim_dataset}. Due to the limited number of such kind of objects in MUSIC, several clusters are in common with all the 10 sub samples. In all the cases, the gNFW profile parameters are consistent ($1\sigma$), see Table \ref{Tab:param_P_nika}, second row, but different to the MUSIC ones in Table \ref{Tab:param_P_highz}. Repeating the same procedure, without constraining the mass range, leads to random samples which have gNFW profile parameters compatible with the MUSIC one, an example of parameters is listed in the third row of Table \ref{Tab:param_P_nika}. The difference may be due to the number of high mass clusters in the NIKA2 twin sample, larger in proportion than the MUSIC one. 

\begin{table*}
\centering
\caption{Best fit parameters and errors to the gNFW pressure profile for the NIKA2 twin sample (first row). The second row shows the same information but for a random selection of MUSIC clusters with the same mass range, redshift and dynamical state than the NIKA2 sample, while in the third row, the random selection is done without any mass constrain. }
\begin{tabular}[t]{cccccc}
\hline
 & $P_0$  &  $c_{500}$  &  $a$  &  $b$  &  $c$  \\ 
\hline
NIKA2 twin sample & $10.21\pm2.55$ & $0.19\pm0.10$ & $0.78\pm0.06$ & $12.13\pm3.10$ & $0.20\pm0.06$ \\
Random sample & $16.89\pm5.37$ & $0.18\pm0.09$ & $0.77\pm0.06$ & $12.92\pm3.10$ &  $0.09\pm0.06$ \\
Random sample (no mass cut) & $3.79\pm0.35$ & $2.04\pm0.07$ & $1.82\pm0.12$ & $3.71\pm0.07$ & $0.56\pm0.03$ \\
\hline
\end{tabular} 
\label{Tab:param_P_nika}
\end{table*}

\begin{figure}
	\includegraphics[width=.45\textwidth]{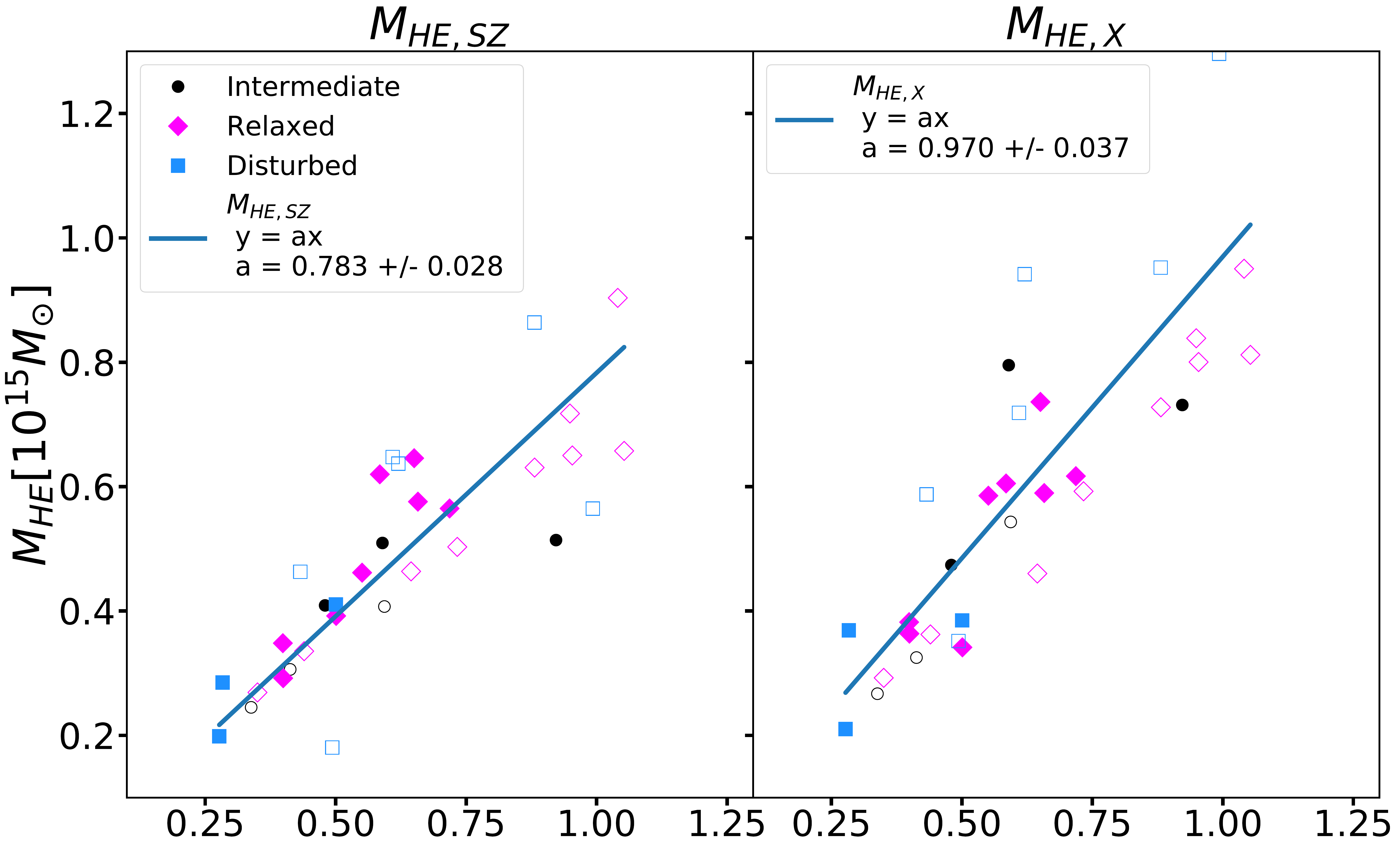}
	\includegraphics[width=.45\textwidth]{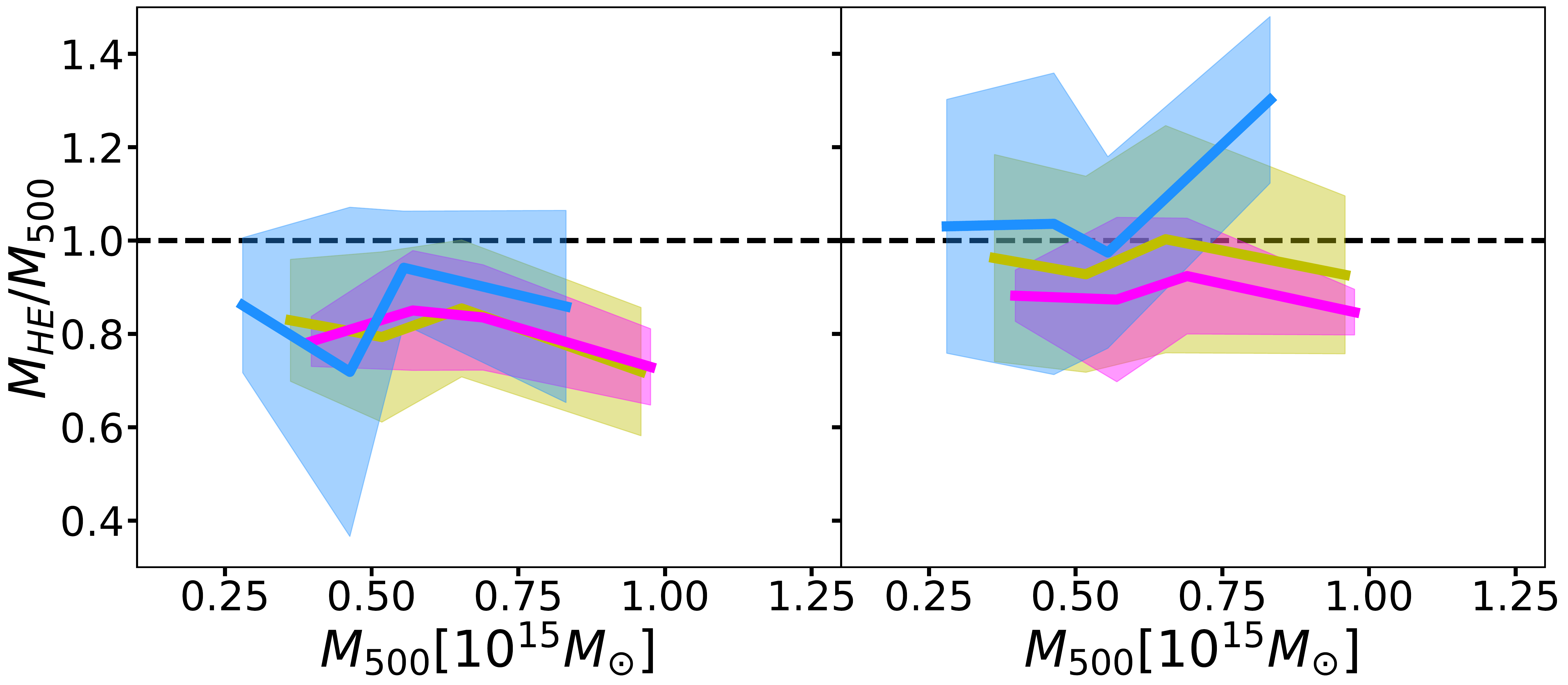}
 	\caption{ Same as Fig. \ref{fig:an_deriv_mhe_m500}, but for the NIKA2 sub sample. In the first row, the clusters at $z=0.54$ are represented as open symbols, while those at $z=0.82$ as solid symbols.}
 \label{fig:NIKA2_an_deriv_mhe_m500}
\end{figure}

In the first row of Fig. \ref{fig:NIKA2_an_deriv_mhe_m500} the HE masses, estimated for the NIKA2 twin sample, are shown as a function of the true cluster mass. These HE masses were estimated using the analytical fitting, for both redshifts and all simulation flavours. In this figure, we just show the results for the AGN simulations, both at $z=0.54$, (open symbols), and $z =0.82$, (solid symbols). Also in this case a fit of the type $M_{\rm HE} = aM_{500}$ was performed, the slopes $a$ are listed in Table \ref{Tab:NIKA2_an_deriv_slopes}. The slopes and their errors on the NIKA2 twin sample fits are higher with respect to the full MUSIC sample, (see Table \ref{Tab:an_deriv_slopes_good}), due to the difference in the mass distributions. In the second row of Fig. \ref{fig:NIKA2_an_deriv_mhe_m500} the ratio $M_{\rm HE}/M_{500}$ is represented as a function of $M_{500}$, also in this case, as in the MUSIC sample, we do not find any bias dependence on the mass.

\begin{table*}
	\caption{The best fit values and their corresponding errors to the parameter $a=1-b$ in the $M_{\rm HE} = aM_{500}$ relation for the NIKA2 twin sample in the two redshift bins and for all the simulation flavours.}
	\begin{tabular}{ccccccc} 
 \textbf{$z$} & \multicolumn{3}{c}{\textbf{Slope $a_{\rm SZ}$}} & \multicolumn{3}{c}{\textbf{Slope $a_{\rm X}$}}\\
      &  AGN   &  CSF   &  NR  &  AGN   &  CSF   &  NR  \\
\hline
0.54+0.82 & $0.783\pm0.028$  & $0.948\pm0.071$  &  $0.775\pm0.034$ &$0.970\pm0.037$   & $0.952\pm0.030$  & $0.972\pm0.050$   \\
\hline
	\end{tabular}
	\label{Tab:NIKA2_an_deriv_slopes}
\end{table*}

\begin{figure}
	\includegraphics[width=.45\textwidth]{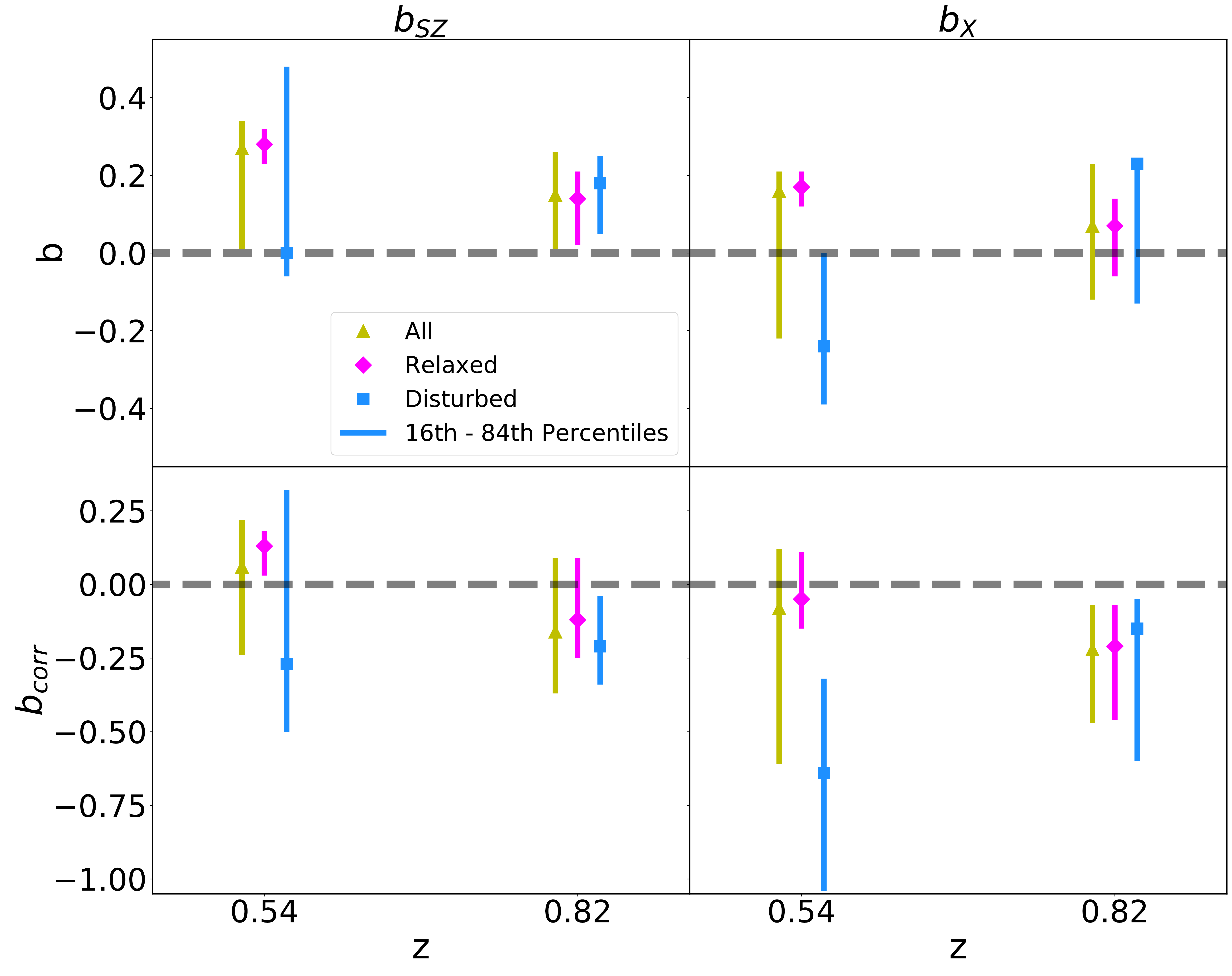}
 	\caption{The redshift evolution of the median values of the SZ and X-ray bias (first row) and the corrected ones, (second row), for the NIKA2 twin sample, using the AGN flavour and the analytical fitting method. The relaxed, disturbed and all clusters cases are represented in magenta diamonds, blue squares and yellow triangles respectively. The error bars represent the 16th and 84th percentile.}
 \label{fig:NIKA2_an_deriv_b}
\end{figure}

The HE mass biases of the NIKA2 twin sample are represented in the first row of Fig. \ref{fig:NIKA2_an_deriv_b} along the redshifts, using the analytical fitting method. They are slightly higher than the correspondent MUSIC cases in Fig. \ref{fig:b_vs_z_model_good}, at reshifts 0.54 and 0.82, but still in agreement with them. Contrary to MUSIC full sample, here the disturbed clusters show a bias close to 0 at $z=0.54$. Large errors make these biases still compatible with the whole sample. We have to report that this unexpected behaviour could be also due to having not excluded the clusters with non reliable fits. However, the disturbed clusters show the same behaviour using the numerical derivative method.

The same problem reflects on the corrected bias. The correction on the biases has been applied to this sample as well, their median values are shown in the second row of Fig. \ref{fig:NIKA2_an_deriv_b}. We see that, at redshift $0.54$, the non-thermal correction works well, giving bias values close to 0, compatible with the MUSIC sample. Instead, at redshift $0.82$ the corrected bias is slightly lower than 0, although still compatible with it within 2 $\sigma$. At $z=0.54$, the correction does not work for the case of disturbed clusters, but it is still in agreement with the MUSIC value. 

\section{Mass bias distributions}
\label{app:distributions}

\begin{figure}
    \includegraphics[width=.45\textwidth]{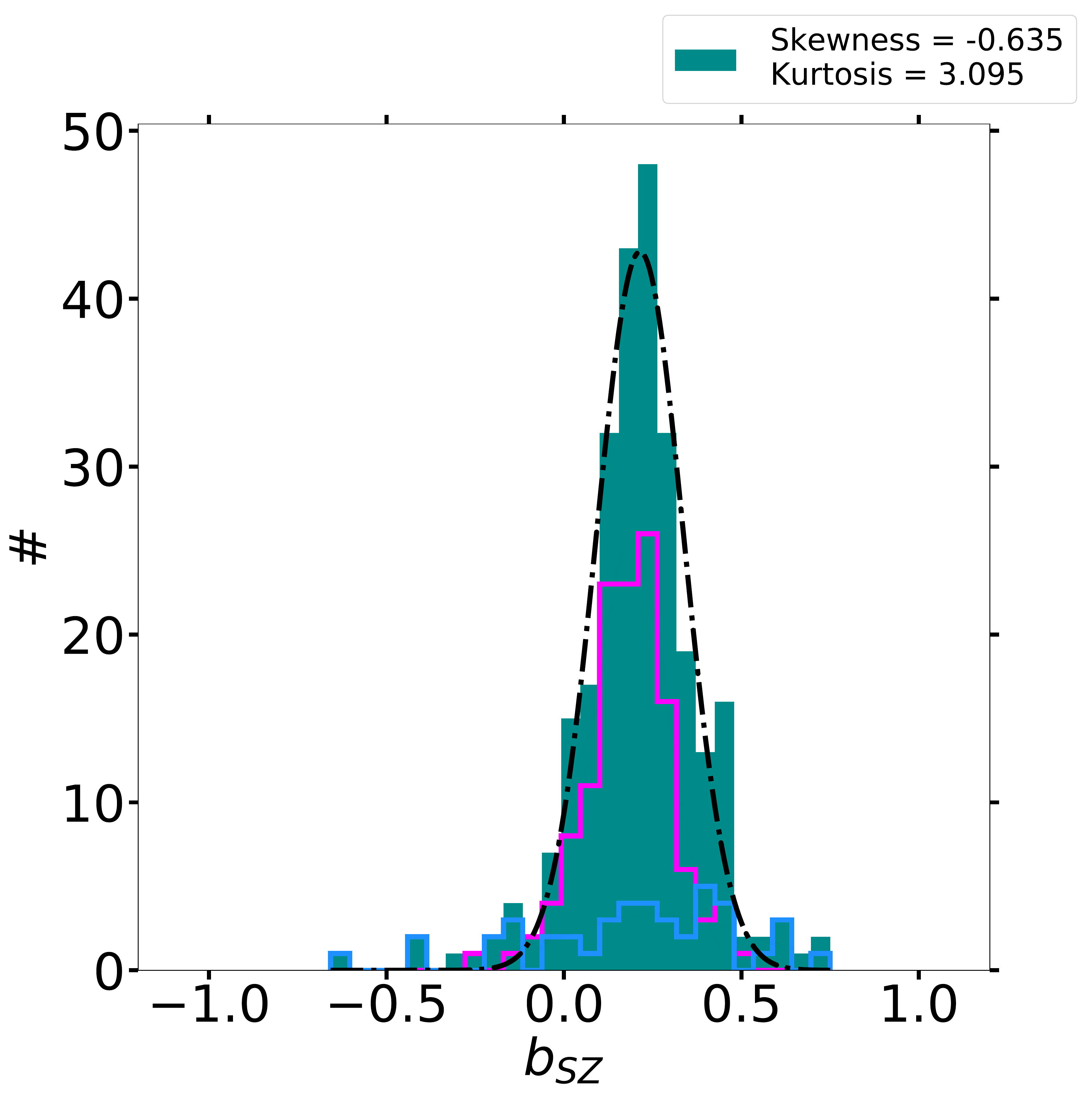}
    \includegraphics[width=.45\textwidth]{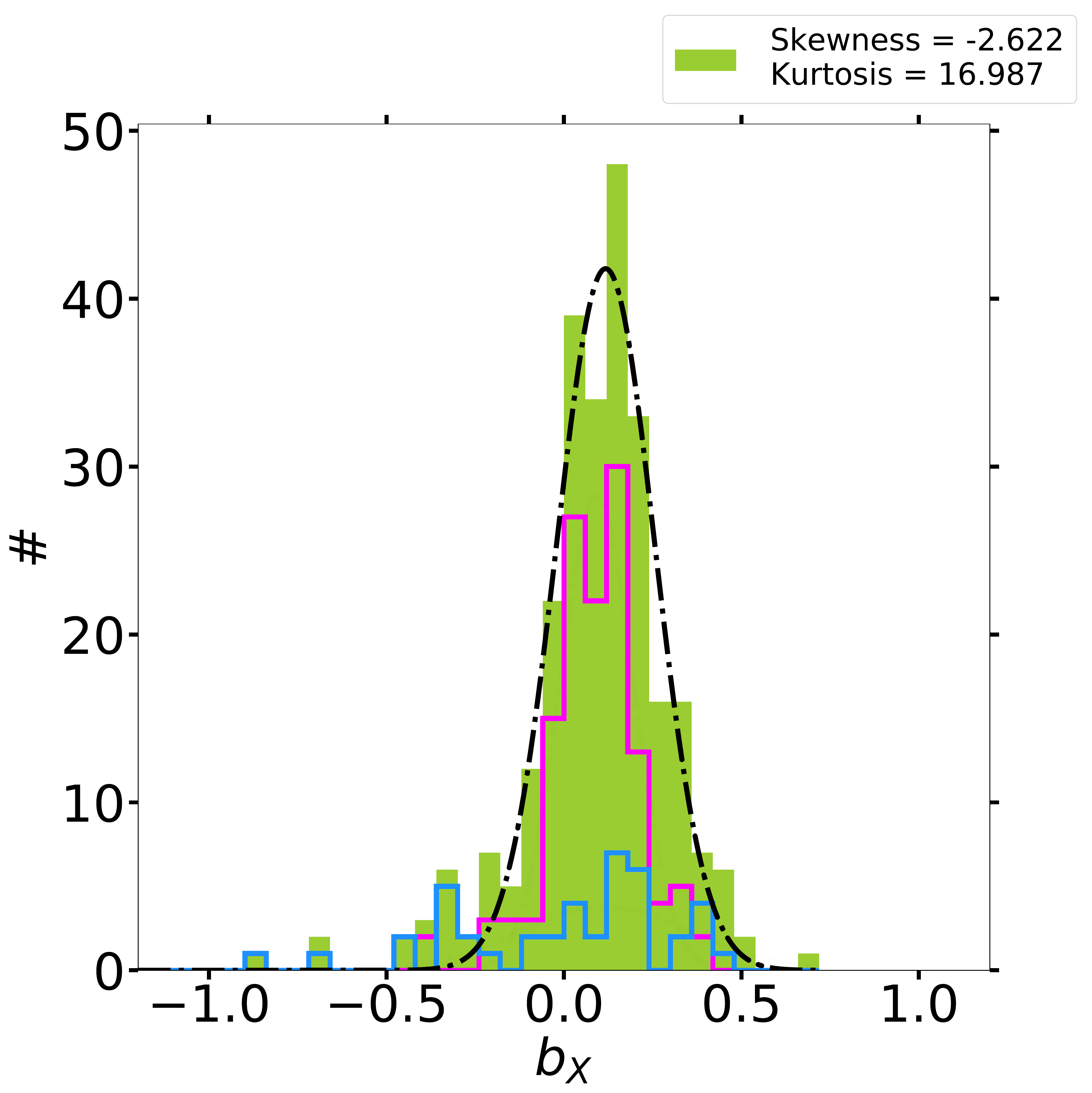}
	\caption{Histograms of the $b_{\rm SZ}$ and $b_{\rm X}$ values corresponding to AGN simulations of the complete MUSIC sample, using the analytical fitting method, for the whole MUSIC sample, without discriminating for the goodness of the fit $\hat{\chi}^2$. In dark cyan (left figure) and light green (right figure) there are the biases of all the clusters, in magenta, for only relaxed and in blue for only disturbed clusters. In the legend, the values of the skewness and the kurtosis are reported.}
	\label{fig:hist_AGN}
\end{figure}

Examples of bias distribution are shown in Fig.\ref{fig:hist_AGN} for the analytical fitting method, without discriminating for the goodness of the fits $\hat{\chi}^2$, for the AGN simulation at $z=0$. In Table \ref{Tab:data}, we report the means and the standard deviations from the data and from a Gaussian fit for both methods. The distributions are not Gaussian, as evident from the skewness and kurtosis values, reported in the legends of Fig.\ref{fig:hist_AGN}. The Anderson test also confirms that the distributions are not Gaussian and thus we will not comment on the mean and standard deviation. Only in a few cases the distributions are actually Gaussian. In the Table the bootstrap errors, $err_{\rm b}$, are listed also. For all cases they are of the order of $10^{-2}$.

\begin{table}
\centering
\caption{The results in this Table refer to the $b_{\rm X}$ and $b_{\rm SZ}$ distributions at redshift $z=0$ and for the AGN runs. The value of the mean $\mu$ and standard deviation $\sigma$ estimated both from the \textit{Data} and from a Gaussian \textit{fit} of the bias distributions. Moreover, the bootstrap error is written. All these informations are reported for both methods, the \textit{Analytical} fitting and the \textit{Numerical} derivative one. }
\begin{tabular}[t]{|c|c|c|c|c|c|c|c|c|}
\hline
  &  &  & \multicolumn{3}{c}{Analytical} & \multicolumn{3}{c}{Numerical} \\
\hline
  &  &  & $\mu$  & $\sigma$ & $err_{\rm b}$ &  $\mu$ & $\sigma$ & $err_{\rm b}$ \\
\hline
All & $b_{\rm X}$ & Data & 0.09 & 0.23 & 0.01 & 0.13 & 0.26 & 0.02\\ 
    &       & Fit & 0.12 & 0.20 &  & 0.13 & 0.22 & \\ 
    & $b_{\rm SZ}$ & Data & 0.21 & 0.18 & 0.01 & 0.19 & 0.26 & 0.02\\ 
    &      & Fit & 0.21 & 0.17 &  & 0.17 & 0.22 & \\ 
 \hline
Rel & $b_{\rm X}$ & Data & 0.08 & 0.13 & 0.01 & 0.09 & 0.15 & 0.01\\ 
    &   & Fit & 0.09 & 0.15 &  & 0.11 & 0.18 & \\ 
 &$b_{\rm SZ}$ & Data & 0.19 & 0.12 & 0.01 & 0.13 & 0.15 & 0.01\\ 
  &        & Fit & 0.19 & 0.15 &  & 0.14 & 0.17 & \\ 
 \hline
Dis & $b_{\rm X}$ & Data & -0.03 & 0.40 & 0.06 & 0.12 & 0.44 & 0.07\\ 
    &   & Fit & 0.12 & 0.34 &  & 0.10 & 0.32 & \\ 
 &$b_{\rm SZ}$ & Data & 0.19 & 0.30 & 0.05 & 0.21 & 0.38 & 0.06\\ 
   &       & Fit & 0.12 & 0.34 &  & 0.11 & 0.37 & \\ 
 \hline

\end{tabular}
\label{Tab:data}
\end{table}

We can say that the two methods to compute the derivative lead to very similar and compatible results, nevertheless it is worth noticing that the biases from the numerical derivative method show a broader distribution. 
Computing the mass bias through the X-ray and SZ equations based on different ICM properties also leads to similar results on the HE bias. However, in the case of the analytical fitting method, $b_{\rm SZ}$ and $b_{\rm X}$ have both negative skewness for all the redshifts, meaning that they have a more pronounced tail on the left side of the distribution, i.e. toward the 0 bias, as found also in \citet{bib:rasia19}. They study the distributions of the $b_{\rm SZ}$ and $b_{\rm X}$, estimated using the gas density $\rho_{\rm g}$ instead of $\mu m_p n_{\rm e}$ in Eq. (\ref{eq:Mhe_P}) and (\ref{eq:Mhe_T}), finding that $b_{\rm SZ}$ shows a larger scatter and that $b_{\rm X}$ is more symmetric than $b_{\rm SZ}$. These results disagree with what we found here. We conclude that the differences are due to the radial gas density profiles considered to estimate $M_{\rm HE}$. In fact, if we use $\rho_g$, we have the same distribution features as  in \citet{bib:rasia19}.

\begin{figure}
 \includegraphics[width=.45\textwidth]{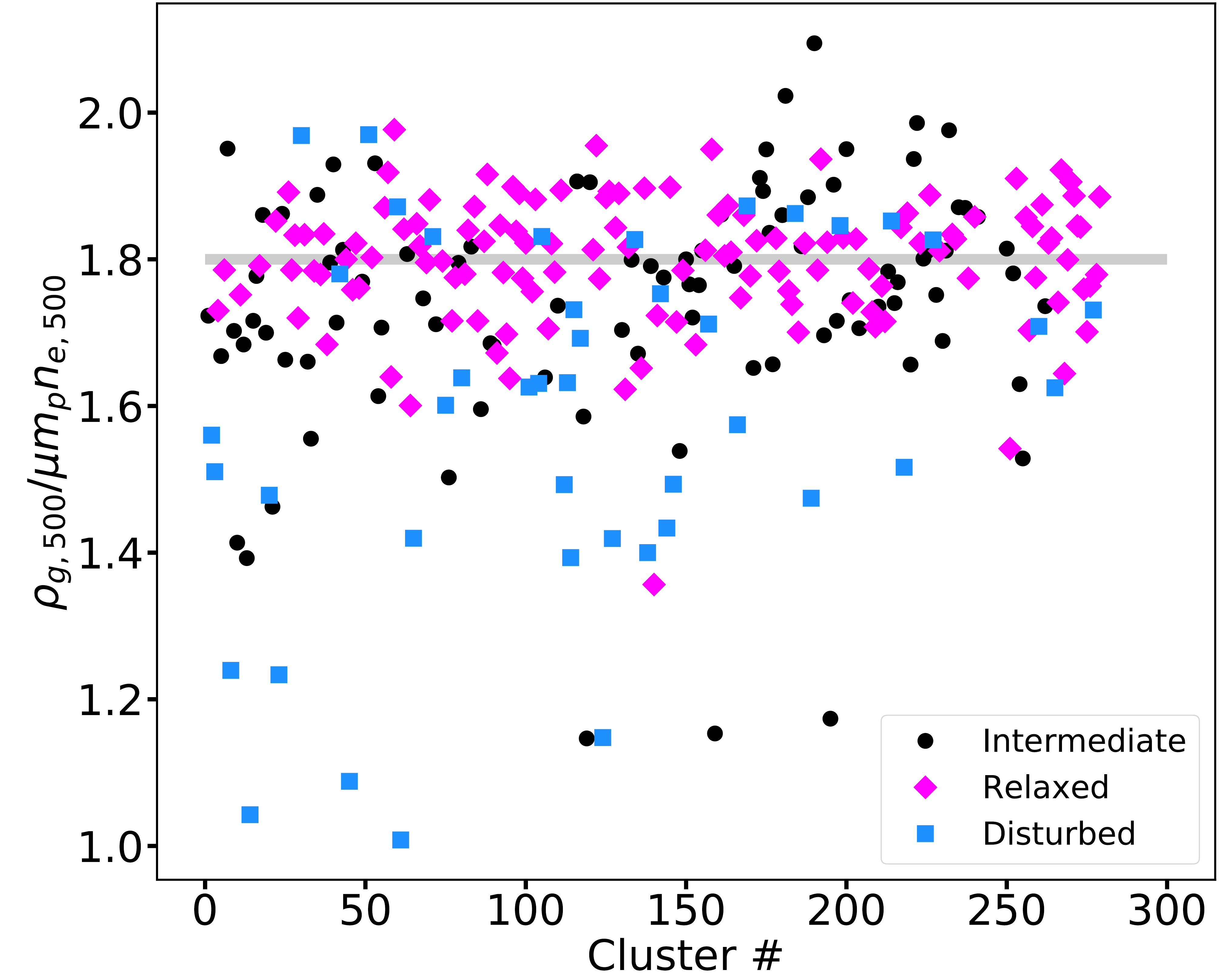}
 \caption{The fraction of the gas density and numerical electron density at $R_{500}$ is represented for each cluster in the AGN flavour at $z = 0$. The relaxed, disturbed and intermediate clusters are represented in magenta diamonds, blue squares and black circles respectively. The grey line is the expected value of 1.8.}
 \label{fig:rho_ne_500}
\end{figure}

There should not be any difference between the results obtained with $\rho_{\rm g}$ and those obtained with $\mu m_p n_{\rm e}$, because they should be interchangeable quantities. However, we find that this relation sometime is not satisfied, and probably for this reason we have different biases for the electron and for the gas density. The gas density $\rho_g$ in the MUSIC simulations is computed as the total gas mass in the spherical shell divided by the shell volume. The numerical electron density (in $cm^{-3}$) is evaluated from the gas density using
\begin{equation}
n_{\rm e}(r) = N_{\rm e} \rho_{\rm g}(r) \left( \frac{1 - Z - Y_{\rm He}}{m_p} \right),
\label{eq:ne}
\end{equation}
$N_{\rm e}$ is the fraction of free electrons per hydrogen particle, $Z$ is the metallicity, $Y_{\rm He}$ is the nuclear helium concentration and $m_p$ is the proton mass \citep{bib:semb}. The numerical electron density takes into account the local distribution of electrons and gas particles, if the gas is completely ionized, then we can relate the electron and gas density using
\begin{equation}
 \rho_{\rm g}(r) = 1.8 \mu m_p n_{\rm e}(r)
\label{eq:ne_approx}
\end{equation}
where $\mu$, the mean molecular weight of electrons, is 0.59 \citep{bib:rev_x}. We find that this approximation does not hold in the outskirts of the cluster, where the contribution of non-thermal motions, like bulk motions or turbulence, starts to be important (see Section \ref{subsec:P_nth}). Moreover, for the same reason, we find that some disturbed and intermediate clusters do not follow this relation either. We show this behaviour in Fig. \ref{fig:rho_ne_500}, where the ratio $\rho_{\rm g} / \mu m_p n_{\rm e}$ at $R_{500}$ for each cluster is represented. From the Figure, we see that this relation is valid mostly for the relaxed clusters, in magenta diamonds. While in Fig. \ref{fig:rho_ne_500} we show only the example of AGN at $z=0$, this happens for all redshifts and flavours. 

\section{Mass bias results}
We present in Table \ref{tab:res_bias}, the HE mass biases for all $z$ and for all flavours, using only the reliable fits from the analytical fitting method.

 \begin{table*}
  \caption{A compilation of the HE mass bias estimates $b_{\rm SZ}$ and $b_{\rm X}$ and their corresponding corrected ones (at the bottom part of the Table) computed at $R_{500}$ for clusters with reliable fits to the analytical profiles in the MUSIC sample, for different redshifts (first column). The biases are listed taking into account $All$ the clusters and only the $Rel$axed or $Dis$turbed ones. The results for the three AGN, CSF and NR simulation flavours are listed as median values with 16th and 84th percentile errors. }
  \label{tab:res_bias}
  \begin{tabular}[t]{c|c|c|c|c|c|c|c}
 \textbf{$z$} & \textbf{Clusters} & \multicolumn{2}{|c|}{\textbf{AGN}} & \multicolumn{2}{|c|}{\textbf{CSF}} & \multicolumn{2}{|c}{\textbf{NR}} \\
      &     & $b_{\rm SZ}$      &  $b_{\rm X}$    &  $b_{\rm SZ}$     &    $b_{\rm X}$  &   $b_{\rm SZ}$   &  $ b_{\rm X}$  \\
\hline
    & All & $0.23^{+0.14}_{-0.09}$ & $0.14^{+0.11}_{-0.13}$ & $0.26^{+0.12}_{-0.1}$ & $0.14^{+0.16}_{-0.11}$ & $0.27^{+0.15}_{-0.11}$ & $0.15^{+0.14}_{-0.13}$ \\
0.0 & Rel & $0.19^{+0.09}_{-0.05}$ & $0.08^{+0.13}_{-0.06}$ & $0.23^{+0.09}_{-0.09}$ & $0.13^{+0.1}_{-0.1}$  & $0.25^{+0.11}_{-0.05}$ & $0.14^{+0.11}_{-0.12}$ \\
    & Dis & $0.29^{+0.1}_{-0.1}$   & $0.14^{+0.09}_{-0.18}$  & $0.32^{+0.28}_{-0.5}$ & $0.12^{+0.23}_{-0.34}$ & $0.26^{+0.22}_{-0.59}$ & $0.21^{+0.06}_{-0.32}$ \\
\hline 
    & All & $0.25^{+0.11}_{-0.11}$ & $0.14^{+0.13}_{-0.13}$ & $0.25^{+0.14}_{-0.12}$ & $0.12^{+0.18}_{-0.14}$  & $0.28^{+0.13}_{-0.11}$ & $0.14^{+0.16}_{-0.12}$ \\
0.11 & Rel& $0.2^{+0.1}_{-0.12}$ &  $0.11^{+0.09}_{-0.08}$  & $0.2^{+0.13}_{-0.08}$ & $0.11^{+0.11}_{-0.09}$ & $0.22^{+0.09}_{-0.05}$ & $0.12^{+0.09}_{-0.09}$\\
    & Dis & $0.41^{+0.05}_{-0.17}$ & $0.18^{+0.22}_{-0.2}$ & $0.35^{+0.2}_{-0.21}$ & $0.14^{+0.23}_{-0.28}$ & $0.4^{+0.14}_{-0.26}$ & $0.2^{+0.18}_{-0.16}$ \\
\hline 
    & All & $0.24^{+0.13}_{-0.09}$ & $0.15^{+0.13}_{-0.12}$ & $0.27^{+0.11}_{-0.12}$ & $0.19^{+0.09}_{-0.16}$ & $0.3^{+0.09}_{-0.12}$ & $0.2^{+0.1}_{-0.16}$ \\
0.33 & Rel& $0.22^{+0.09}_{-0.07}$ & $0.12^{+0.11}_{-0.11}$ & $0.23^{+0.1}_{-0.09}$ & $0.17^{+0.08}_{-0.14}$ & $0.29^{+0.07}_{-0.07}$ & $0.18^{+0.08}_{-0.14}$  \\
    & Dis & $0.35^{+0.1}_{-0.16}$ &  $0.26^{+0.08}_{-0.28}$ & $0.37^{+0.14}_{-0.15}$ & $0.18^{+0.13}_{-0.31}$ & $0.41^{+0.12}_{-0.21}$ & $0.21^{+0.1}_{-0.2}$ \\
\hline 
    & All & $0.26^{+0.13}_{-0.12}$ & $0.13^{+0.14}_{-0.13}$ & $0.28^{+0.1}_{-0.14}$ & $0.15^{+0.12}_{-0.14}$  & $0.31^{+0.09}_{-0.1}$ & $0.19^{+0.14}_{-0.17}$ \\
0.43 & Rel& $0.22^{+0.11}_{-0.08}$ & $0.12^{+0.06}_{-0.09}$ & $0.27^{+0.08}_{-0.12}$ & $0.14^{+0.12}_{-0.12}$ & $0.3^{+0.06}_{-0.09}$ & $0.14^{+0.13}_{-0.12}$ \\
    & Dis & $0.27^{+0.2}_{-0.25}$ & $0.13^{+0.23}_{-0.23}$  & $0.35^{+0.14}_{-0.31}$ & $0.15^{+0.2}_{-0.4}$ & $0.36^{+0.1}_{-0.32}$ & $0.23^{+0.19}_{-0.31}$ \\
\hline 
    & All & $0.24^{+0.1}_{-0.14}$ & $0.12^{+0.12}_{-0.14}$ & $0.28^{+0.11}_{-0.14}$ & $0.12^{+0.15}_{-0.15}$ & $0.3^{+0.11}_{-0.15}$ & $0.18^{+0.12}_{-0.14}$ \\
0.54 & Rel& $0.23^{+0.09}_{-0.11}$ & $0.11^{+0.1}_{-0.08}$ & $0.26^{+0.07}_{-0.1}$ & $0.12^{+0.13}_{-0.11}$ & $0.27^{+0.06}_{-0.12}$ & $0.16^{+0.09}_{-0.12}$ \\
    & Dis & $0.33^{+0.11}_{-0.22}$ & $0.14^{+0.15}_{-0.28}$ & $0.33^{+0.2}_{-0.22}$ & $0.12^{+0.18}_{-0.29}$ & $0.41^{+0.07}_{-0.1}$ & $0.21^{+0.1}_{-0.16}$ \\
\hline 
    & All & $0.25^{+0.13}_{-0.16}$ & $0.14^{+0.15}_{-0.17}$ & $0.31^{+0.11}_{-0.12}$ & $0.16^{+0.12}_{-0.17}$ & $0.28^{+0.13}_{-0.11}$ & $0.18^{+0.18}_{-0.18}$ \\
0.67 & Rel& $0.24^{+0.09}_{-0.12}$ & $0.13^{+0.1}_{-0.13}$  & $0.28^{+0.07}_{-0.08}$ & $0.13^{+0.09}_{-0.12}$ & $0.26^{+0.09}_{-0.09}$ & $0.15^{+0.09}_{-0.15}$  \\
    & Dis & $0.25^{+0.18}_{-0.25}$ & $0.11^{+0.19}_{-0.28}$ & $0.38^{+0.11}_{-0.22}$ & $0.16^{+0.2}_{-0.23}$ & $0.33^{+0.14}_{-0.15}$ & $0.23^{+0.2}_{-0.2}$  \\
\hline
    & All & $0.25^{+0.11}_{-0.13}$ & $0.14^{+0.12}_{-0.17}$ & $0.27^{+0.13}_{-0.15}$ & $0.15^{+0.16}_{-0.15}$ & $0.29^{+0.09}_{-0.14}$ & $0.17^{+0.15}_{-0.15}$ \\
0.82 & Rel& $0.25^{+0.09}_{-0.09}$ & $0.12^{+0.1}_{-0.15}$  & $0.26^{+0.09}_{-0.1}$ & $0.15^{+0.13}_{-0.13}$ & $0.31^{+0.05}_{-0.11}$ & $0.17^{+0.11}_{-0.11}$  \\
    & Dis & $0.27^{+0.13}_{-0.28}$ & $0.17^{+0.25}_{-0.26}$ & $0.3^{+0.23}_{-0.33}$ & $0.13^{+0.29}_{-0.19}$ & $0.16^{+0.24}_{-0.41}$ & $0.07^{+0.29}_{-0.2}$\\
\hline 
      &     & $b_{\rm SZ, corr}$      &  $b_{\rm X, corr}$    &  $b_{\rm SZ, corr}$     &    $b_{\rm X, corr}$  &   $b_{\rm SZ, corr}$   &  $ b_{\rm X, corr}$  \\
\hline
    & All & $0.12^{+0.15}_{-0.15}$ & $-0.02^{+0.19}_{-0.19}$ & $0.14^{+0.19}_{-0.23}$ & $0.0^{+0.19}_{-0.24}$ & $0.14^{+0.2}_{-0.19}$ & $0.01^{+0.18}_{-0.23}$ \\
0.0 & Rel & $0.1^{+0.12}_{-0.11}$ & $-0.03^{+0.14}_{-0.11}$ & $0.1^{+0.13}_{-0.2}$ & $-0.01^{+0.15}_{-0.21}$ & $0.11^{+0.15}_{-0.13}$ & $0.01^{+0.15}_{-0.16}$ \\
    & Dis & $0.14^{+0.19}_{-0.38}$ & $-0.2^{+0.17}_{-0.26}$ & $0.18^{+0.35}_{-0.44}$ & $-0.15^{+0.32}_{-0.35}$ & $-0.16^{+0.54}_{-0.52}$ & $-0.2^{+0.24}_{-0.35}$\\
\hline 
    & All & $0.15^{+0.12}_{-0.2}$ & $-0.02^{+0.18}_{-0.17}$ & $0.12^{+0.2}_{-0.19}$ & $-0.01^{+0.18}_{-0.25}$  & $0.16^{+0.14}_{-0.22}$ & $-0.0^{+0.17}_{-0.22}$ \\
0.11 & Rel & $0.1^{+0.11}_{-0.19}$ & $-0.03^{+0.15}_{-0.14}$ & $0.09^{+0.17}_{-0.14}$ & $-0.01^{+0.12}_{-0.19}$ & $0.13^{+0.1}_{-0.21}$ & $0.0^{+0.14}_{-0.14}$  \\
    & Dis & $0.16^{+0.21}_{-0.19}$ & $-0.12^{+0.51}_{-0.37}$ & $0.12^{+0.34}_{-0.41}$ & $-0.13^{+0.39}_{-0.57}$ & $0.25^{+0.1}_{-0.48}$ & $-0.05^{+0.21}_{-0.38}$ \\
\hline 
    & All & $0.11^{+0.15}_{-0.2}$ & $-0.04^{+0.19}_{-0.26}$ & $0.09^{+0.24}_{-0.24}$ & $-0.03^{+0.21}_{-0.28}$ & $0.15^{+0.2}_{-0.21}$ & $-0.03^{+0.18}_{-0.33}$ \\
0.33 & Rel & $0.06^{+0.15}_{-0.22}$ & $-0.07^{+0.17}_{-0.21}$ & $0.02^{+0.2}_{-0.15}$ & $-0.06^{+0.19}_{-0.25}$ & $0.11^{+0.21}_{-0.17}$ & $-0.05^{+0.2}_{-0.26}$\\
    & Dis & $0.18^{+0.18}_{-0.31}$ & $0.0^{+0.26}_{-0.37}$ & $0.25^{+0.23}_{-0.41}$ & $-0.06^{+0.21}_{-0.57}$  & $0.23^{+0.24}_{-0.33}$ & $-0.01^{+0.17}_{-0.45}$\\
\hline 
    & All & $0.14^{+0.18}_{-0.2}$ & $-0.01^{+0.23}_{-0.24}$ & $0.14^{+0.15}_{-0.28}$ & $-0.01^{+0.19}_{-0.27}$ & $0.16^{+0.13}_{-0.23}$ & $-0.01^{+0.21}_{-0.33}$\\
0.43 & Rel & $0.1^{+0.16}_{-0.17}$ & $-0.03^{+0.15}_{-0.22}$ & $0.13^{+0.14}_{-0.26}$ & $-0.04^{+0.2}_{-0.25}$ & $0.15^{+0.1}_{-0.17}$ & $-0.06^{+0.18}_{-0.29}$ \\
    & Dis & $0.13^{+0.29}_{-0.38}$ & $-0.1^{+0.43}_{-0.25}$  & $0.18^{+0.22}_{-0.6}$ & $-0.08^{+0.23}_{-0.6}$ & $0.2^{19}_{-0.48}$ & $0.01^{+0.26}_{-0.53}$ \\
\hline 
    & All & $0.12^{+0.16}_{-0.24}$ & $-0.05^{+0.2}_{-0.28}$ & $0.12^{+0.18}_{-0.23}$ & $-0.1^{+0.22}_{-0.31}$ & $0.14^{+0.16}_{-0.18}$ & $-0.02^{+0.16}_{-0.36}$ \\
0.54 & Rel & $0.12^{+0.09}_{-0.15}$ & $-0.02^{+0.16}_{-0.25}$ & $0.12^{+0.1}_{-0.18}$ & $-0.09^{+0.19}_{-0.19}$ & $0.08^{+0.13}_{-0.13}$ & $-0.04^{+0.16}_{-0.25}$\\
    & Dis & $0.12^{+0.23}_{-0.35}$ & $-0.13^{+0.24}_{-0.41}$  & $0.12^{+0.36}_{-0.39}$ & $-0.22^{+0.34}_{-0.44}$ & $0.2^{+0.11}_{-0.13}$ & $-0.09^{+0.19}_{-0.3}$ \\
\hline 
    & All & $0.08^{+0.23}_{-0.28}$ & $-0.04^{+0.22}_{-0.27}$ & $0.13^{+0.22}_{-0.25}$ & $-0.09^{+0.26}_{-0.24}$ & $0.09^{+0.25}_{-0.26}$ & $-0.09^{+0.27}_{-0.34}$ \\
0.67 & Rel & $0.06^{+0.21}_{-0.23}$ & $-0.04^{+0.13}_{-0.22}$ & $0.1^{+0.12}_{-0.18}$ & $-0.09^{+0.14}_{-0.22}$ & $0.1^{+0.15}_{-0.22}$ & $-0.12^{+0.18}_{-0.21}
$\\
    & Dis & $0.02^{+0.36}_{-0.33}$ & $-0.19^{+0.34}_{-0.33}$ & $0.16^{+0.31}_{-0.27}$ & $-0.14^{+0.41}_{-0.28}$ & $0.06^{+0.24}_{-0.35}$ & $-0.16^{+0.43}_{-0.29}$ \\
\hline
    & All & $0.11^{+0.21}_{-0.25}$ & $-0.06^{+0.29}_{-0.23}$ & $0.08^{+0.22}_{-0.25}$ & $-0.06^{+0.23}_{-0.31}$ & $0.07^{+0.17}_{-0.3}$ & $-0.09^{+0.18}_{-0.33}$ \\
0.82 & Rel & $0.09^{+0.15}_{-0.19}$ & $-0.07^{+0.12}_{-0.22}$ & $0.07^{+0.12}_{-0.18}$ & $-0.06^{+0.14}_{-0.22}$ & $0.08^{+0.14}_{-0.18}$ & $-0.07^{+0.12}_{-0.25}$ \\
    & Dis & $0.19^{+0.23}_{-0.56}$ & $0.01^{+0.42}_{-0.29}$ & $0.05^{+0.44}_{-0.48}$ & $-0.2^{+0.59}_{-0.29}$ & $-0.13^{+0.52}_{-0.56}$ & $-0.22^{+0.38}_{-0.4}$ \\
\hline 
 \end{tabular} 
 \end{table*}

\end{document}